\documentclass[extra,onecolumn]{gji}
\usepackage[fleqn]{amsmath}
\usepackage{amsfonts,amssymb}
\usepackage{timet,epsfig,wrapfig,psfrag}
\usepackage{natbib}
\usepackage{symbol}
\usepackage{color}
\usepackage[normalem]{ulem}
\usepackage{booktabs}
\usepackage{multirow} 
\usepackage{pdfcolmk} 

\begin{document}

\bibliographystyle{chicago}


\title[Borehole fibre-optic seismology inside the Northeast Greenland Ice Stream]
{Borehole fibre-optic seismology inside the Northeast Greenland Ice Stream}

\author[Andreas Fichtner et al.]{
\parbox{\linewidth}{Andreas Fichtner$^1$, Coen Hofstede$^2$,  Lars Gebraad$^1$,  Andrea Zunino$^1$, Dimitri Zigone$^3$ and Olaf Eisen$^{2,3,4}$}\\ \\
$^1$ Department of Earth Sciences, ETH Zurich, Switzerland\\
$^2$ Alfred Wegener Institute, Helmholtz Centre for Polar and Marine Research, Bremerhaven, Germany\\
$^3$ Universit\'{e} de Strasbourg/CNRS, Institut Terre et Environnement de Strasbourg, France\\
$^4$ Deptartment of Geosciences, University of Bremen, Germany}

\maketitle


\begin{summary}
Ice streams are major contributors to ice sheet mass loss and sea level rise.  Effects of their dynamic behaviour are  imprinted into seismic properties, such as wave speeds and anisotropy.  Here we present results from the first Distributed Acoustic Sensing (DAS) experiment in a deep ice-core borehole in the onset region of the Northeast Greenland Ice Stream.  A series of active surface sources produced clear recordings of the P and S wavefield, including internal reflections, along a 1500 m long fibre-optic cable that was lowered into the borehole.  The combination of nonlinear traveltime tomography with a firn model constrained by multi-mode surface wave data, allows us to invert for P and S wave speeds with depth-dependent uncertainties on the order of only 10 m$/$s, and vertical resolution of 20--70 m. The wave speed model in conjunction with the regularly spaced DAS data enable a straightforward separation of internal upward reflections followed by a reverse-time migration that provides a detailed reflectivity image of the ice.  While the differences between P and S wave speeds hint at anisotropy related to crystal orientation fabric, the reflectivity image seems to carry a pronounced climatic imprint caused by rapid variations in grain size.  Currently, resolution is not limited by the DAS channel spacing. Instead, the maximum frequency of body waves below $\sim$200 Hz, low signal-to-noise ratio caused by poor coupling,  and systematic errors produced by the ray approximation, appear to be the leading-order issues.  Among these, only the latter has a simple existing solution in the form of full-waveform inversion. Improving signal bandwidth and quality, however, will likely require a significantly larger effort in terms of both sensing equipment and logistics.
\end{summary}


\section{Introduction}\label{S:Introduction}

Ice streams are gravitationally-driven motion within ice sheets and a major contributor to their total mass balance.  Greenlandic ice stream discharge amounts to roughly 500 Gt$/$a, thereby turning the Greenland Ice Sheet into the largest single contributor to current sea level rise \citep{King_2020}.  Discovered rather recently by SAR imagery \citep{Fahnestock_1993}, the Northeast Greenland Ice Stream (NEGIS) accounts for $\sim$12 \% of these 500 Gt$/$a, making it the most voluminous active ice stream in Greenland \citep[e.g.,][]{Rignot_2012,Khan_2014}.

The unambiguously observable acceleration of ice discharge from Greenland's marine-terminating outlet glaciers \citep[e.g.,][]{Mouginot_2019,Mankoff_2019,King_2020,Khan_2022} attaches considerable relevance to numerical ice sheet models that aim to predict the effect of climate warming and the resulting consequences for sea level rise and human society \citep{Church_2013}.  The accuracy of these models is limited by our knowledge about the boundary conditions and the rheology of ice streams.  A particularly important contributor to rheology is crystal orientation fabric (COF). On the one hand, COF is part of an internal feedback loop: it controls the macroscopic anisotropy of ice and the resulting flow pattern, which, in turn, modifies the COF \citep[e.g.,][]{Alley_1992,Pettit_2007,Martin_2009}.  On the other hand, COF is affected by external factors,  such as changes in dust load associated with a climatic transitions, volcanic ash deposition or basal micro-particles \citep[e.g.,][]{Diprinzio_2005,Samyn_2005,Durand_2007}. 

COF and other rheologic properties may be measured directly in ice cores \citep[e.g.,][]{Bennett_1968,Faria_2013,Faria_2014}. However, deep ice core drilling is logistically demanding, usually of long duration and expensive. Furthermore, the \emph{in situ} azimuthal orientation of the ice crystals in the horizontal plane is generally lost during the retrieval of the core.  Indirect geophysical methods, e.g., seismic and radar) constitute a potentially attractive alternative to direct measurements because COF may lead to both anisotropic wave propagation \citep[e.g.,][]{Picotti_2015,Gerber_2023} and reflections from regions where the fabric changes rapidly compared to the wavelength \citep[e.g.,][]{Diez_2015a,Diez_2015b}.  However, as radar observations operate at very small offsets between source and receiver, they can usually only provide information about the horizontal anisotropy, not the full fabric. Only passive or active seismic methods may provide the full COF. Successful estimates of depth-dependent ice fabric based on surface seismic reflection data my be found, for instance, in \cite{Bentley_1972}, \cite{Blankenship_1987} and \cite{Horgan_2011}.

The emergence of fibre-optic sensing technologies, and of Distributed Acoustic Sensing (DAS) in particular, opens new opportunities for the study of icy materials with elastic waves. In addition to dense spatial sampling at metre scale, DAS offers a large bandwidth, ranging from mHz to kHz under favorable coupling conditions \citep{Lindsey_2020,Paitz_2021,Bernauer_2021}. The ease of trenching a fibre-optic cable in snow or ice makes DAS especially attractive for seismic studies on glaciers and ice sheets \citep[e.g.,][]{Walter_2020,Klaasen_2021,Klaasen_2022,Hudson_2021,Fichtner_2022b,Fichtner_2023,Zhou_2023}. 

To infer ice stream properties with high vertical resolution, DAS measurements in deep boreholes would be desirable.  Without the need to recover an ice core, they could provide nearly space-continuous access to physical properties at depth,  and probe a larger and potentially more representative volume.  In seismic reservoir monitoring and exploration, pioneering DAS applications in boreholes date back around one decade \citep[e.g.,][]{Mateeva_2013,Daley_2013,Mateeva_2014,Daley_2014}. In contrast, similar experiments on ice sheets are more recent \citep{Booth_2020,Brisbourne_2021}.

\begin{figure}
\begin{center}
\noindent\includegraphics[width=1.0\textwidth]{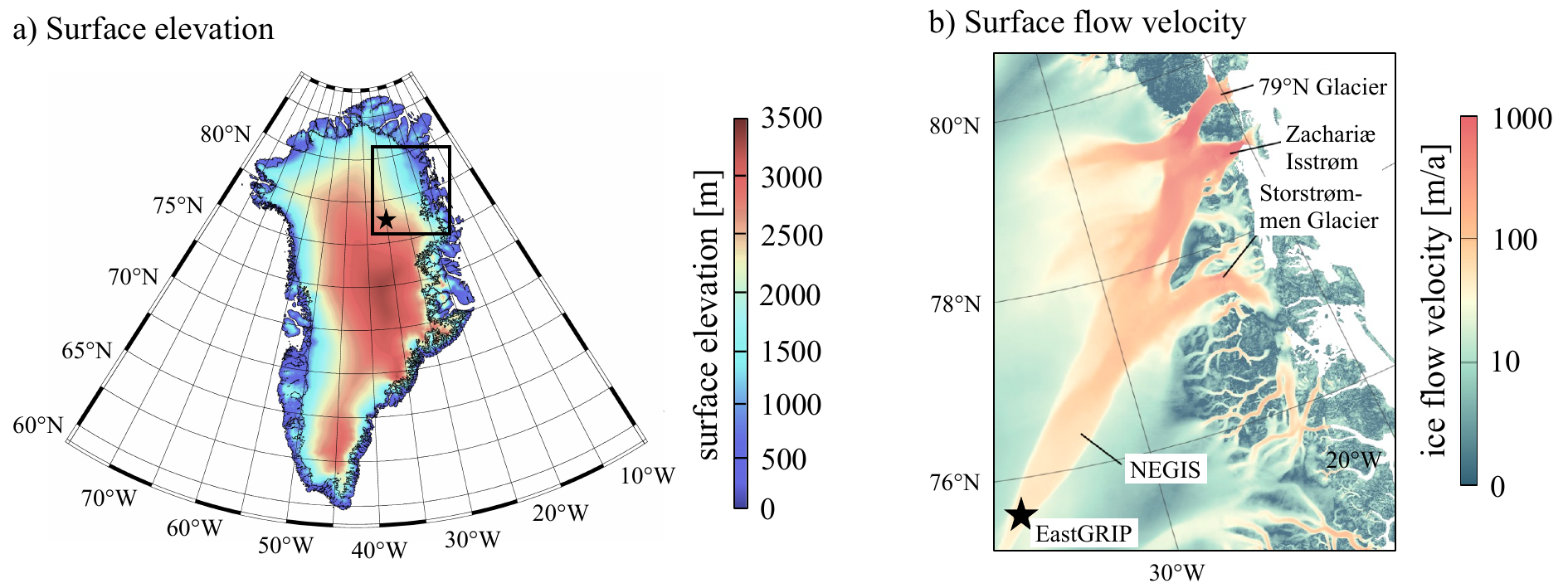}
\caption{Geographic setting. a) Surface elevation of the Greenland Ice Sheet, derived from CryoSat-2 data. The location of the EastGRIP drill site is marked by the black star. The area shown in panel b) is outlined by the black rectangle.  Figure modified from \cite{Helm_2014}. b) Surface flow velocity of NEGIS \citep{Joughin_2018} with its outlet glaciers.}
\label{F:location}
\end{center}
\end{figure}

Here we present results from the first DAS deployment in a deep ice-core borehole in an active ice stream.  As part of the East Greenland Ice-Core Project (EastGRIP), the site is located in the onset region of the NEGIS, where surface flow velocities are around 50 m$/$a  (Fig. \ref{F:location}).  Reaching to 1500 m depth, the DAS recordings contain clearly distinguishable direct and reflected P and S waves originating from explosive sources at the surface.  They provide a comprehensive picture of wave speed variations and seismic reflectivity within the ice stream. 

The focus of this work is on phenomenological and methodological aspects that may guide future fibre-optic seismology projects in deep ice-core boreholes.  These aspects include (i) the nature of the seismic wavefield excited by active sources, (ii) the depth- and frequency-dependent characteristics of noise, (iii) the nonlinear inversion for P and S wave speed profiles, (iv) reflectivity imaging, and (v) the achievable resolution in relation to the major aleatoric and epistemic uncertainties. A detailed glaciological interpretation of the inversion results will later be based on these analyses, and will involve the integration of independent data, e.g., from ice-core crystallography and radar soundings.

\section{Experimental setup}\label{S:setup}

On 9 August 2022, we lowered a Solifos BRUfield\texttrademark\, fiber-optic cable by hand into the EastGRIP borehole. The light-weight cable with a diameter of 38 mm contained four optical fibers and had a density of $\sim$1150 kg$/$m$^3$.  Its short- and long-term tensile strengths were 1200 N and 650 N, respectively. 

At the time of the experiment,  the borehole had reached a depth of $\sim$2420 m by electro-mechanical ice-core deep drilling; around 240 m above the ice-bed interface, estimated from radio-echo sounding \citep{Vallelonga_2014}.  In deeper boreholes, drilling fluid is required to avoid borehole closure from overburden pressure. To avoid leakage of the fluid into the pore space of the firn column, a borehole casing is commonly used, reaching from the top of the borehole to below the firn-ice transition. The EastGRIP borehole is filled with a mix of two thirds ESTISOL\texttrademark\,  240 and one third COASOL\texttrademark\, \citep{Sheldon_2014} to $\sim$70 m below the surface, approximately the depth where the borehole casing ends.  At an average temperature of around -20$^\circ$C,  the drill fluid has a density of $\sim$940 kg$/$m$^3$ and a kinematic viscosity of $\sim$27 mm$^2/$s, i.e., 27 times that of water at room temperature.

Owing to the high viscosity of the drill fluid,  the cable would not sink by its own weight at an acceptable, or even noticeable, speed.  An additional mass of 2 kg, attached to the cable end, was required in order to achieve a sinking speed of $\sim$0.2 m$/$s. To ensure that the tensile strength of the cable would not be exceeded, we tested every few hundred metres if the cable could still be pulled up by hand. In the absence of experience with similar experiments and trying to be conservative, we decided to stop at 1500 m depth, where we estimated the force required to retrieve the cable at $\sim$250 N.

Thanks to the slight inclination of the borehole of $\sim$3$^\circ$ on average, the cable was likely resting on the borehole wall instead of hanging freely within the drill fluid. Consequently, the cable was mechanically coupled to the ice of the borehole wall, thereby making a significant contribution to the data quality, which we will discuss in more detail in section \ref{S:phenomenology}.

\begin{figure}
\begin{center}
\noindent\includegraphics[width=1.0\textwidth]{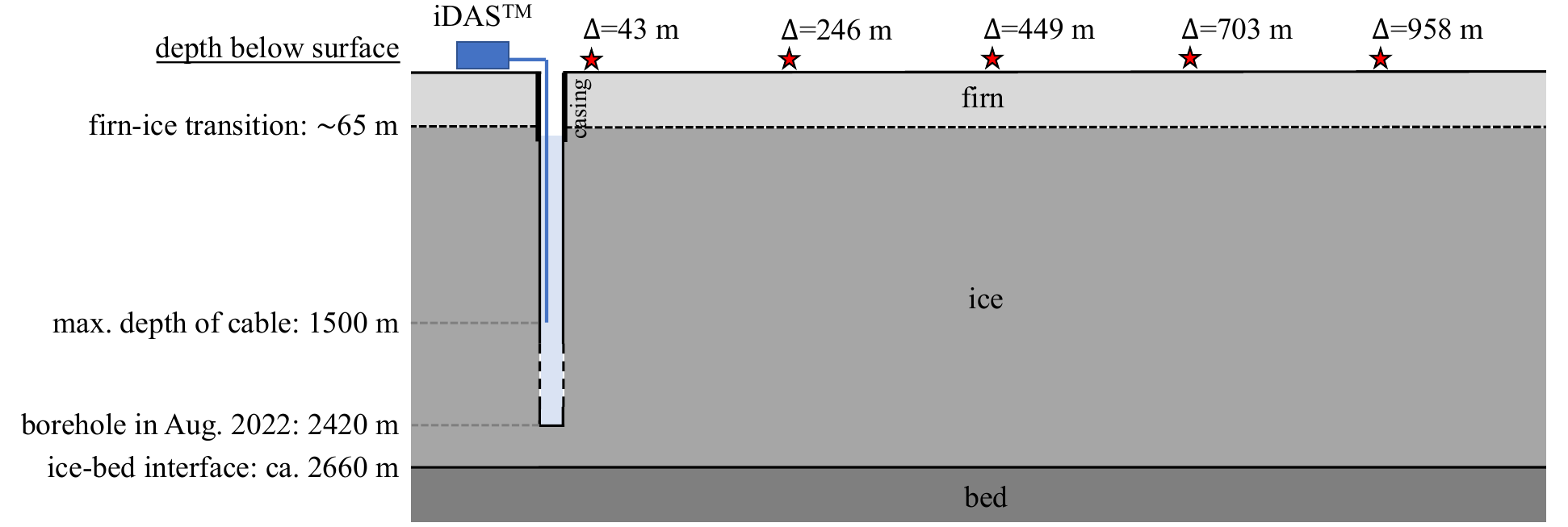}
\caption{Schematic, not-to-scale, illustration of the experimental setup.  The DAS cable, shown as blue line, reaches a maximum depth of 1500 m inside the 2620 m deep borehole. Shots, marked by red stars, were fired at distances $\Delta$ from the borehole every $\sim$200 to $\sim$250 m along a straight line.  Drill core and seismological data independently locate the firn-ice transition around 65 m depth \citep{Vallelonga_2014,Fichtner_2023}. The ice-bed interface is approximately 2660 m below the surface \citep{Vallelonga_2014}.}
\label{F:setup}
\end{center}
\end{figure}

To one of the fibers we connected a Silixa iDAS\texttrademark\, interrogator with a gauge length of 10 m.  We set the channel spacing and the sampling frequency to 2 m and 1 kHz, respectively.  Recording during the experiment was continuous in order to avoid the technical complication of remote triggering. As illustrated in Fig. \ref{F:setup}, we fired surface shots consisting of 200 g PETN in the form of a 20 m detonation cord, doubled to 10 m length and placed parallel with the shot line every $\sim$200 to $\sim$250 m along a straight line with an azimuth of -16.7$^\circ$ from true north (i.e., NNW), starting at the borehole. 

\section{Phenomenology}\label{S:phenomenology}

\begin{figure}
\begin{center}
\noindent\includegraphics[width=1.0\textwidth]{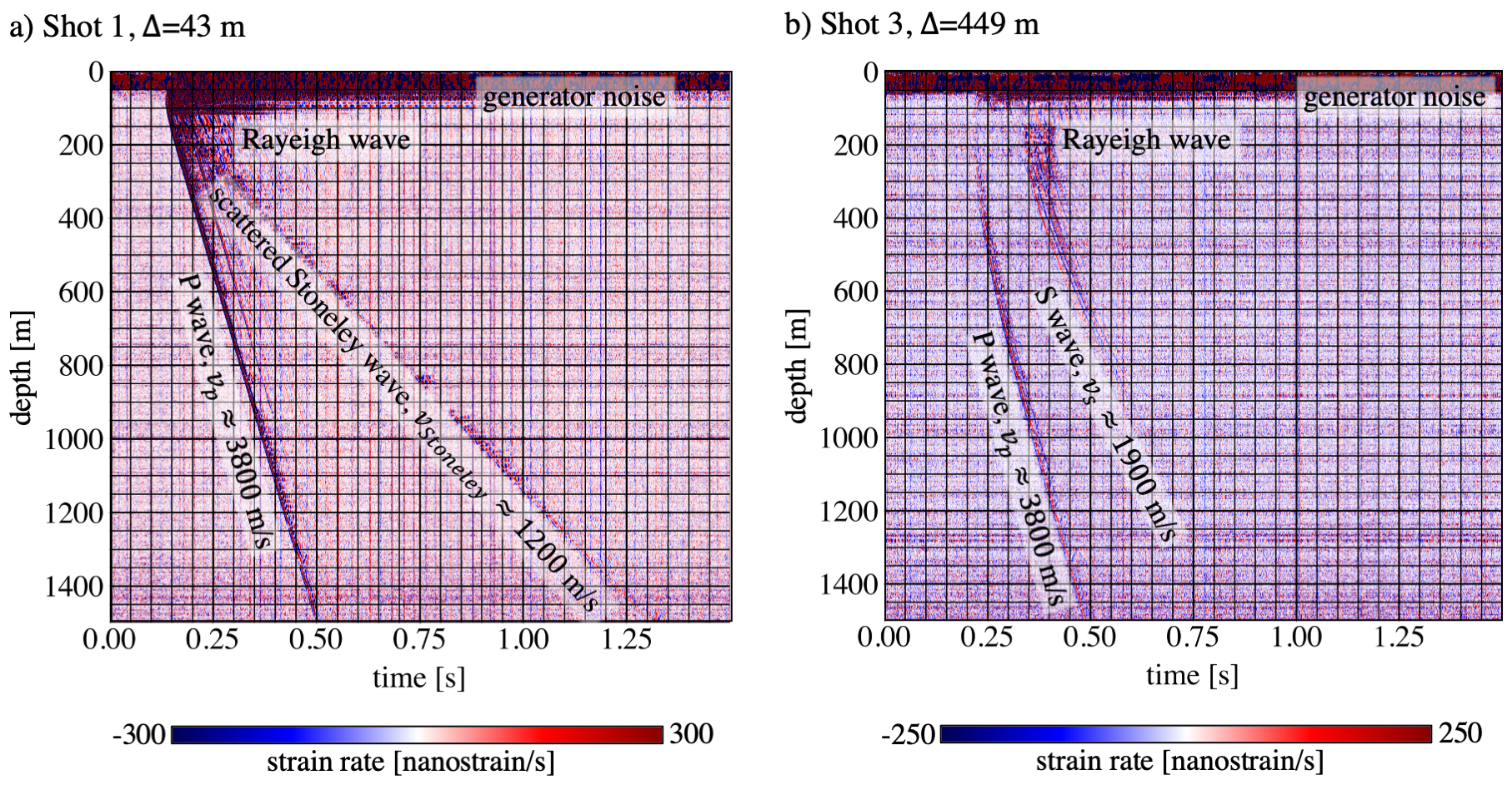}
\caption{Raw strain rate recordings for shot 1 at a distance of $\Delta$=43 m from the borehole (a) and shot 3 at $\Delta$=449 m (b). Major constituents of the wavefield, labelled in the figure, include near-surface generator noise, Rayleigh waves, P and S waves, as well as scattered Stoneley waves for shot 1, fired close to the borehole.}
\label{F:phenomenology}
\end{center}
\end{figure}

As illustrated in Fig. \ref{F:phenomenology}, the most prominent signal in the DAS strain rate recordings is anthropogenic noise from the Diesel generator in the EastGRIP camp.  Despite being concentrated around a frequency of $\sim$30 Hz, its large amplitude of up to 20'000 nanostrain/s completely overwhelms the explosion-generated near-surface signals,  with amplitudes that are typically around two orders of magnitude smaller. Thin vertical stripes in Fig. \ref{F:phenomenology} are most likely high-frequency optical noise that affects all channels simultaneously.

Below $\sim$50 m depth, the amplitude of the generator noise diminishes quickly, revealing a dispersed Rayleigh wave train related to the active shot, which dominates the wavefield to around 100--200 m depth.  For shot 1, fired at a distance of $\Delta$=43 m from the borehole,  direct P waves with a velocity of $v_p \approx$3800 m$/$s are clearly visible down to the end of the cable at 1500 m depth, as shown in Fig. \ref{F:phenomenology}a.  Mostly due to the downward-directed radiation pattern of a predominantly vertical single force \citep{Kennett_2001,Aki_Richards_2002},  direct P wave amplitudes decay quickly for the other shots with increasing distance from the borehole. Beyond shot 5 at $\Delta$=958 m, P waves cannot be detected with confidence.

Also as a consequence of the radiation pattern of an explosive source, S waves with a velocity of $v_s\approx$1900 m$/$s can be recorded only at larger offsets.  They start to be clearly visible in shot 3 at $\Delta$=449 m, shown in Fig. \ref{F:phenomenology}b.  Beyond shot 5 at $\Delta$=958 m, visco-elastic attenuation and geometric spreading reduce the S wave amplitudes below the noise level, which is around 50 nanostrain$/$s for frequencies between $\sim$50--200 Hz, where body waves are most prominent.

\begin{figure}
\begin{center}
\noindent\includegraphics[width=1.0\textwidth]{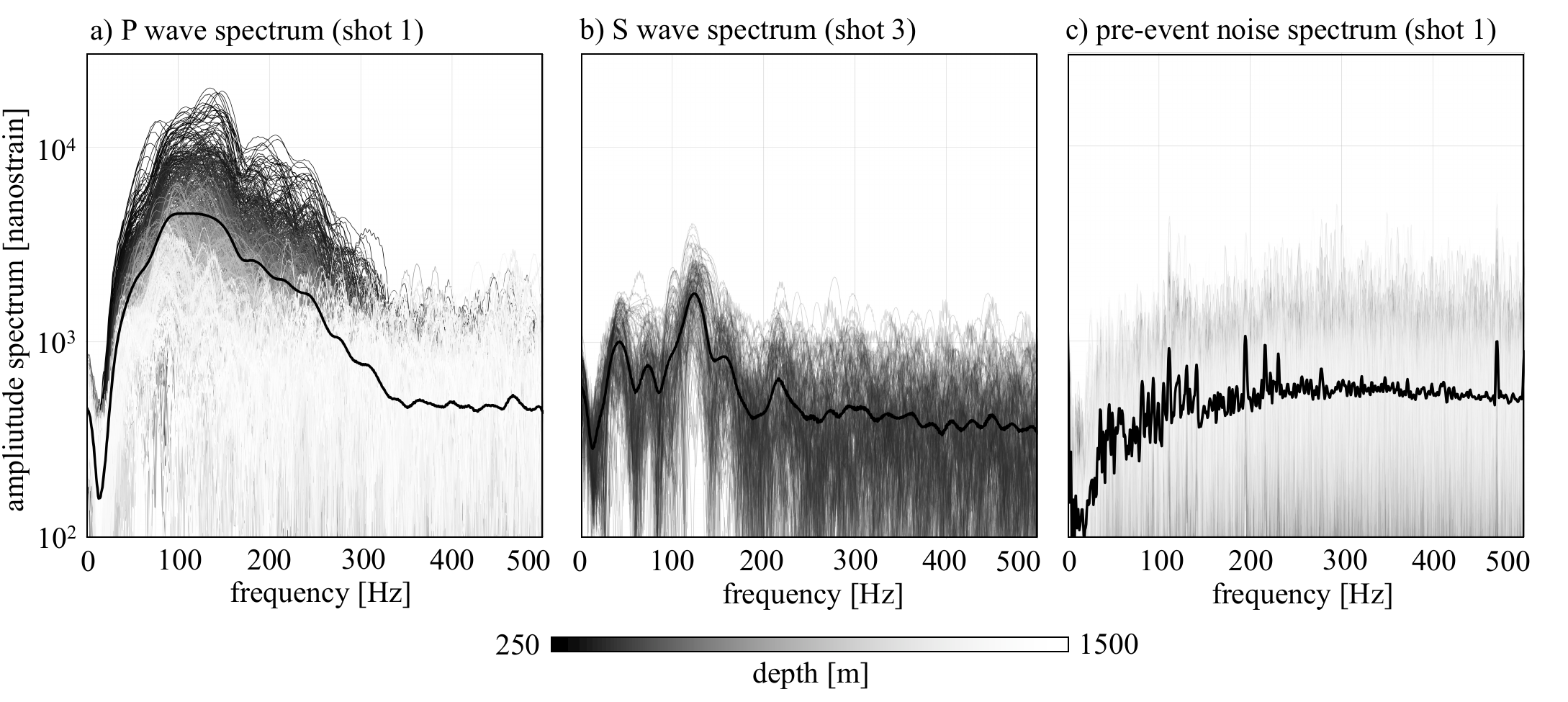}
\caption{Amplitude spectra of P waves (a), S waves (b) and noise recorded prior to the P wave arrival (c). The grey scale encodes the depth of an individual trace, starting at 250 m depth, where anthropogenic (generator) noise has largely decayed below a noticeable level. The thick black curve marks the average amplitude spectrum over all traces. For the S waves in panel 3, the maximum depth is 800 m, below which the S wave cannot be detected with reasonable confidence.}
\label{F:spectra}
\end{center}
\end{figure}

The detailed amplitude spectra of the P waves from shot 1 are displayed in Fig. \ref{F:spectra}a as a function of depth.  Beyond 300 Hz, the P wave amplitude is comparable to the background noise, shown in Fig. \ref{F:spectra}. This frequency band limitation justifies the application of a lowpass filter with 300 Hz cutoff and the application of a sinc interpolation that converts the discretely sampled data into an effectively continuous signal. As explained in section \ref{S:traveltimes}, the latter is important for the estimation of body wave traveltime shifts with an accuracy that is higher than the original sampling rate of 1 kHz.

Fig. \ref{F:spectra}b shows S wave spectra from shot 3. As expected from the time-space domain representation in Fig. \ref{F:phenomenology}b, S waves rise less prominently above the noise level than P waves. Lower amplitudes limit the exploitable frequency range to  $<$150 Hz.  The spectral characteristics of the noise, recorded prior to the P wave arrival and displayed in Fig. \ref{F:spectra}c, are nearly independent of both depth and frequency beyond $\sim$150 Hz. This suggests that the major noise source is the instrument itself. It follows that the instrumental and not the ambient/anthropogenic noise floor controls the detectability of the active-source signals.

\begin{figure}
\begin{center}
\noindent\includegraphics[width=1.0\textwidth]{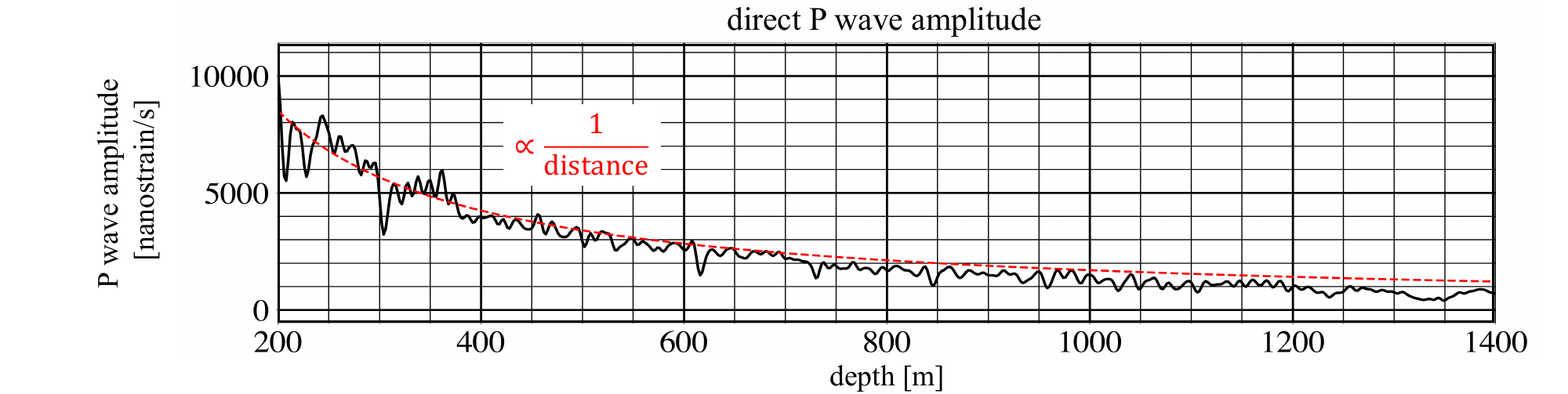}
\caption{The depth-dependent amplitude of the direct P wave from shot 1 closely follows the theoretical geometric spreading of $1/$distance in a homogeneous medium, which is plotted in red for comparison. }
\label{F:amplitudes}
\end{center}
\end{figure}

Fig. \ref{F:amplitudes} illustrates the depth dependence of the unfiltered direct P wave amplitude, estimated as an integral of the signal power over the duration of the P wavelet, $A(z)=[\int \epsilon^2(z,t) dt]^{1/2}$, where $\epsilon(z,t)$ is the strain rate of the P wavelet at depth $z$ for shot 1, fired next to the borehole. 
From 200 m downwards,  where the wavefield is dominated by body waves, the amplitude closely follows the theoretical geometric spreading of $1/$distance in a homogeneous medium.  This is expected well below the firn-ice transition at $\sim$65 m depth \citep{Vallelonga_2014, Fichtner_2023}, where the nearly linear shape of the P wave front suggests only minor variations in P wave speed. Most importantly, the P wave amplitude decay proportional to $1/$distance indicates that coupling of the cable to the surrounding ice did not vary strongly with depth.

\begin{figure}
\begin{center}
\noindent\includegraphics[width=0.9\textwidth]{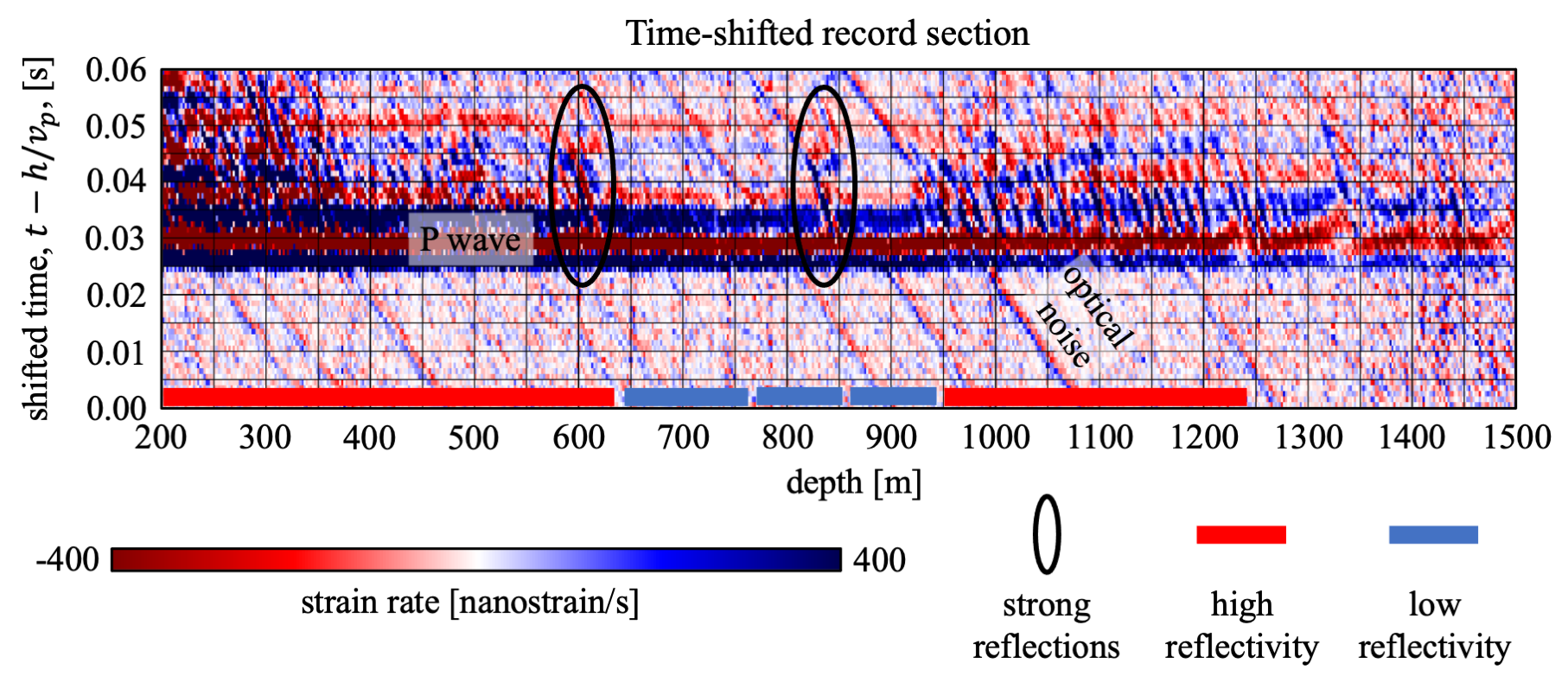}
\caption{Time-shifted section of shot 1 fired close to the borehole at $\Delta$=43 m. Each trace is shifted by $h/v_p$, where $h$ is depth and $v_p$=3800 m$/$s is an approximate average P wave speed. The section is centred around the direct P wave arrival.  Depth intervals where strong reflected waves are visible are marked by horizontal bars.  Blue bars mark depth intervals where strong reflections are essentially absent. Black ellipses indicate easily visible individual reflections.  Optical noise has an apparent propagation speed of exactly $v_p$=3800 m$/$s and corresponds to the purely vertical striping in Fig. \ref{F:phenomenology}.}
\label{F:reduced}
\end{center}
\end{figure}

In \ref{F:reduced}, easily distinguishable upward reflections are visible around 620, 850 and 1050 m depth.  The reflections at 620 and 1050 m separate a depth interval where reflections are rare from depth intervals where reflections abound, thereby suggesting strong vertical variations in reflectivity.  Internal reflections are likely the reason for the direct P wave amplitude decay in Fig. \ref{F:amplitudes}, which is slightly faster than $1/$distance below $\sim$600 m depth. While the easily recognisable reflections enable straightforward reflectivity imaging, they also preclude estimations of the visco-elastic quality factor $Q$ by amplitude ratio methods, because the observed energy loss is not plausibly dominated by visco-elastic attenuation. 

In addition to the P wave, the recording of shot 1 also contains a wave front propagating at a plausible Stoneley wave speed of $\sim$1200 m$/$s.  The presence of an actual Stoneley wave would be surprising because the vertically oriented DAS cable has close to zero sensitivity to horizontally polarised wave motion.  Furthermore, the shape of this wave front is not coherent with depth, and its amplitude varies strongly.  In fact, the largest amplitudes occur at depths where strong P wave reflections can be observed.  This correlation suggests that we rather observe scattered waves excited by the interaction of the Stoneley wave with reflectivity changes in the ice, instead of the Stoneley wave itself.

\section{Traveltime data and inversion}\label{S:traveltimes}

\subsection{P and S wave traveltime observations}

The coherence of P and S waveforms as a function of depth permits the estimation of traveltimes using a matched-filter or template-matching approach \citep{Turin_1960}.  As templates $w(t)$ we use stacks of time-shifted P and S waveforms, with the optimal time shift determined by the maximum of the correlation between neighbouring traces.  This template-matching procedure defines the traveltime of a band-limited (finite-frequency) wave.

To reduce the influence of noise, we limit both the template and the traveltime estimations to traces with a signal-to-noise ratio (SNR) above 3.  Furthermore, we only include traveltimes when the correlation coefficient of the respective trace with the template $w(t)$ is above 0.7. This conservative restriction ensures that the template was actually detected in the noise-contaminated recording with high confidence. Since tight time constraints in the field did not allow us to properly time the shots, absolute traveltimes are only known up to an additive constant.  

\begin{figure}
\begin{center}
\noindent\includegraphics[width=1.0\textwidth]{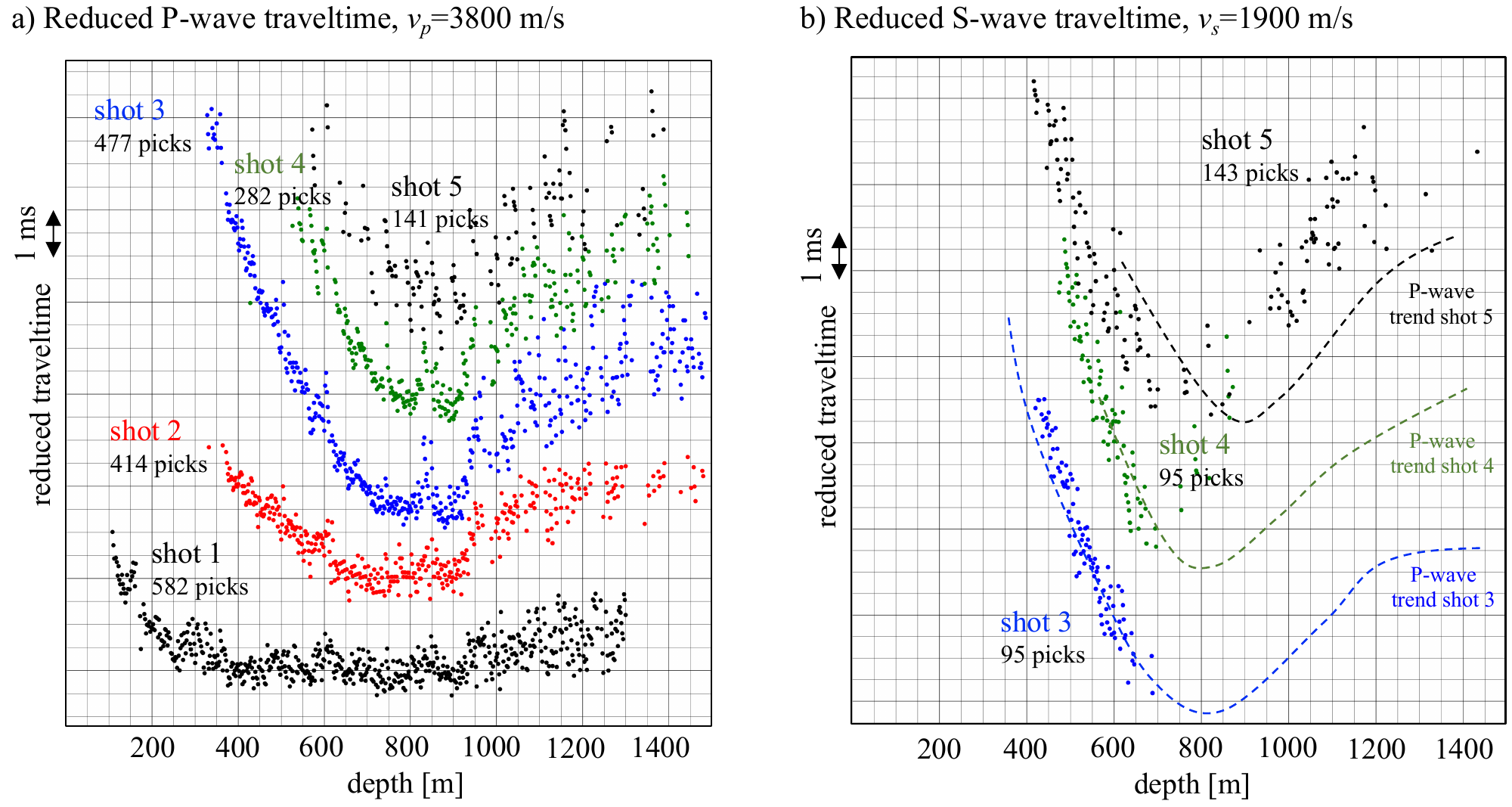}
\caption{Traveltime measurements.  a) Reduced P wave traveltime, $t_0 - \sqrt{h^2 + \Delta^2}/v_p^0$, with a P wave reduction speed of $v_p^0=3800$ m$/$s.  The actual origin time $t_0$ for the different shots is unknown. Specific values for $t_0$ are selected here merely for the purpose of producing a useful summary plot.  b) Reduced S wave traveltimes, similar to panel a), with an S-wave reduction speed of $v_s^0=1900$ m$/$s.  For comparison, the P-wave traveltime trends from panel (a) are superimposed in the form of dashed lines for shots 3, 4 and 5.}
\label{F:traveltimes}
\end{center}
\end{figure}

Fig. \ref{F:traveltimes} summarises the 1896 P and 333 S wave measurements in the form of reduced traveltimes $t_0 - \sqrt{h^2+\Delta^2}/v_{p,s}^0$, where $h$ is depth, $v_p^0$=3800 m$/$s and $v_s^0$=1900 m$/$s. The actual origin time $t_0$ is unknown and set to a value that is convenient for visualisation.  In addition to smaller details with length scales below 100 m, the reduced P wave traveltimes in Fig. \ref{F:traveltimes} show a broad pattern of early arrivals centred around 800 m, suggesting higher than average P wave speed above that depth.  Reduced S wave traveltimes, in contrast, show a different behaviour,  with a minimum roughly between 600--700 m followed by a faster increase, especially for shot 5.  Hence, already a first visual inspection indicates that $v_p$ and $v_s$ behave differently as a function of depth.

Since wave speed variations in ice are expected to be small (in the percent range), measurement uncertainties must be considered with some care. Traveltime uncertainties arising from the presence of quasi-random noise can be estimated by repeatedly measuring the cross-correlation time shift between the template, $w(t)$, and a noise-contaminated version, $w(t)+n(t)$, that plays the role of an artificial data trace. For the noise, $n(t)$, we use 10,000 randomly selected snippets of actual data noise, recorded prior to the P wave arrival.  

\begin{figure}
\begin{center}
\noindent\includegraphics[width=1.0\textwidth]{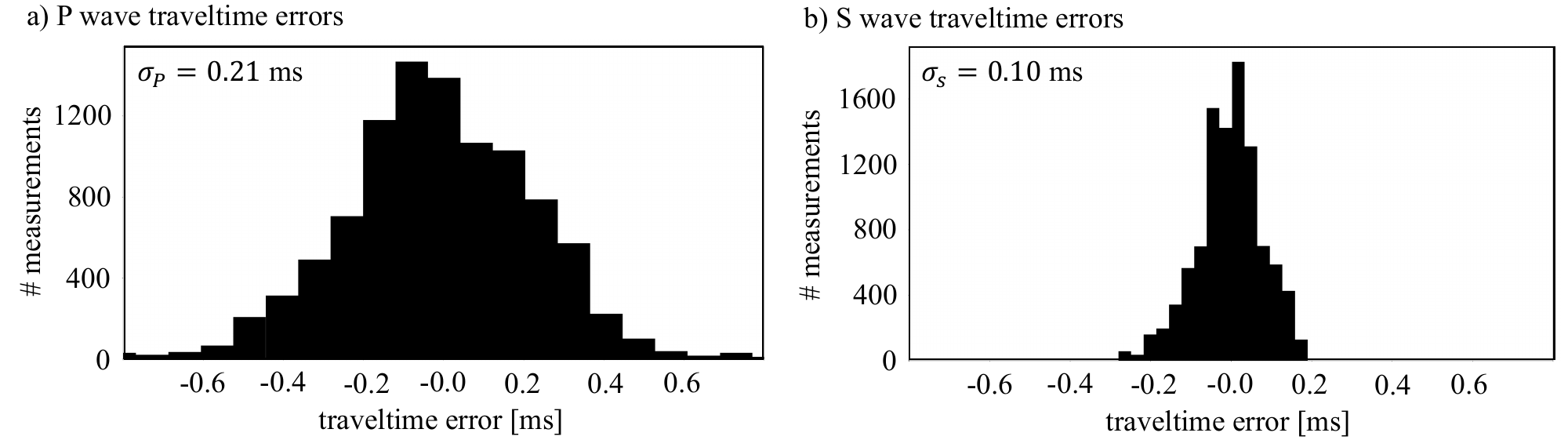}
\caption{Statistical estimation of traveltime measurement uncertainties. a) Histogram of 10,000 random realisations of P wave traveltime estimates between the P wave template from shot 1, $w(t)$, and its noise-contaminated version, $w(t)+n(t)$.  The random noise snippet $n(t)$ is scaled to produce an average SNR of $\sim 3.5$. The standard deviation of the errors, taken as the measurement uncertainty, is $\sigma_P=0.21$ ms. b) The same as in panel (a) but for the S wave template from shot 3. The standard deviation is $\sigma_S=0.10$ ms. Again, the noise snippets are scaled to produce an SNR of $\sim$3.5.}
\label{F:measurement_errors}
\end{center}
\end{figure}

A histogram of the resulting traveltimes for the P wavelet from shot 1 is shown in Fig. \ref{F:measurement_errors}a. The noise snippets were scaled to yield an average SNR of $\sim$3.5.  As expected for noise from a multitude of source, the distribution is unimodal and approximately Gaussian, thereby justifying the use of an $L_2$ misfit for traveltime inversion. The standard deviation of the errors, taken as the measurement uncertainty, is $\sigma_P$=0.21 ms.  Fig. \ref{F:measurement_errors}b shows the result of the uncertainty estimation for the S wavelet from shot 3, again with a scaled noise snippet that produces an SNR of $\sim$3.5. As can be seen in Fig. \ref{F:phenomenology}, the S wavelet has a significantly longer duration than the P wavelet. This leads to a more accurate traveltime measurement, with a standard deviation of $\sigma_S$=0.10 ms.  

\subsection{Nonlinear traveltime inversion}\label{SS:inversion}

In order to avoid a mismatch between the traveltime measurement technique and the inversion method, the finite-frequency definition of traveltimes from section \ref{S:traveltimes} must in principle be matched by full-waveform modelling and an inversion procedure that employs the corresponding finite-frequency sensitivity kernels \citep[e.g.,][]{Luo_Schuster_1991,Marquering_1999,Dahlen_2000}.  Because the computational cost would be prohibitive,  we limit ourselves to forward modelling and inversion based on geometric ray theory. The consequence of this pragmatic simplification is a systematic error that is typically on the order of $\sim$10 \% of the measured traveltime differences for media with wave speed variations in the percent range \citep[e.g.,][]{Baig_2004b,Chaves_2021}.  Conservatively honouring the hardly quantifiable modelling error, we continue with a traveltime uncertainty of 1 ms for both P and S waves.

\begin{figure}
\begin{center}
\noindent\includegraphics[width=0.6\textwidth]{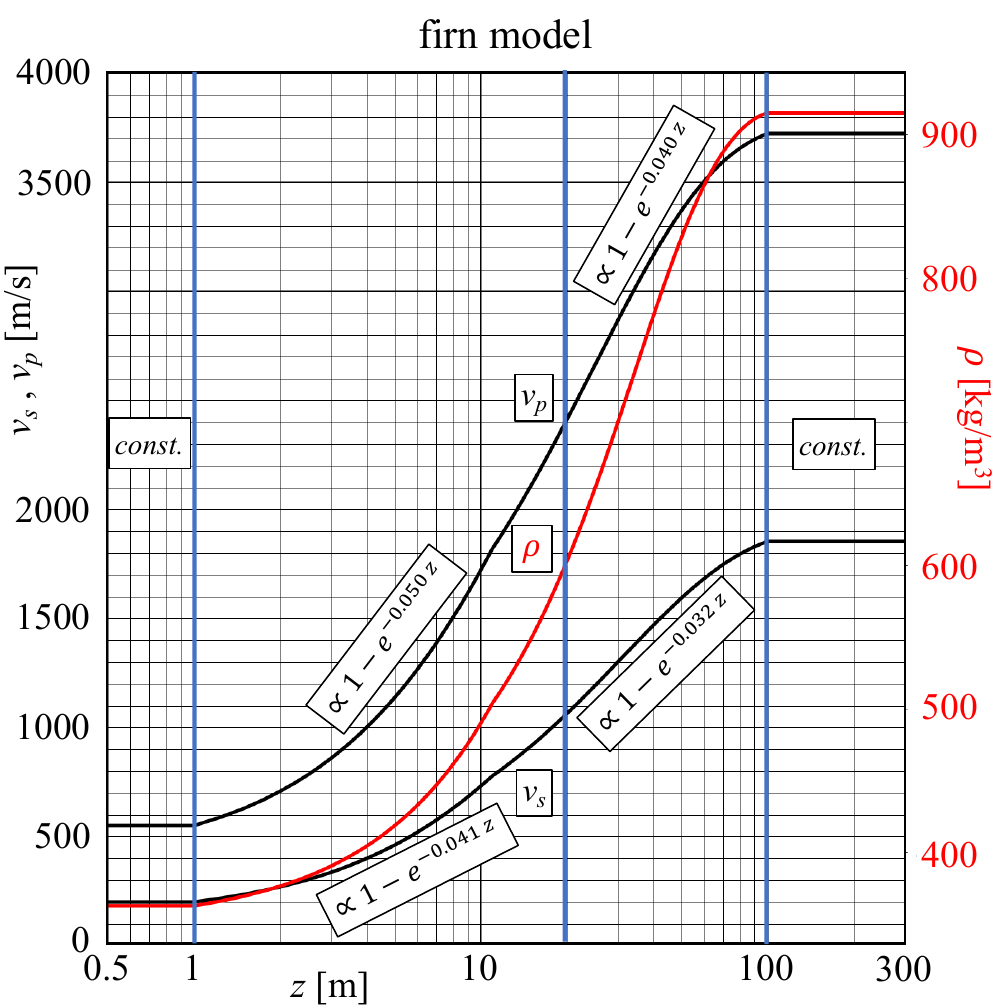}
\caption{A priori P and S wave speed models (black) constrained by multi-mode Rayleigh and P wave mode data collected during an airplane landing near the EastGRIP camp \citep{Fichtner_2023}.  Functional forms in the intervals 0--1 m, 1--11 m, 11--100 m and $>$100 m are indicated for seismic wave speeds. Though not used in the traveltime inversion, the density profile (red) calculated from $v_p$ \citep{Kohnen_1972} is shown for a more complete characterisation of the firn layer.}
\label{F:firn}
\end{center}
\end{figure}

Since the presence of noise combined with the radiation pattern effect do not permit reliable measurements of P and S wave traveltimes near the surface, the upper few hundred metres must be constrained independently prior to a traveltime inversion.  For this, we harness information from multi-mode Rayleigh and P wave modes observed at EastGRIP during the landing of a cargo airplane \citep{Fichtner_2023}.  The firn model that explains the first three Rayleigh modes and the first two P wave modes to within their observational errors is displayed in Fig. \ref{F:firn}.  Possibly due to enhanced compaction by human activity around the EastGRIP camp, the first metre is characterised by constant $v_p$ and $v_s$.  This is followed by an exponential behaviour, typical for firn layers \citep[e.g.,][]{Brockamp_1967,Kohnen_1973}, down to 100 m depth. Below 100 m,  where the modal data gradually loose resolving power, $v_p$ and $v_s$ are initially set to 3720 m$/$s and 1840 m$/$s, respectively.  

\begin{figure}
\begin{center}
\noindent\includegraphics[width=1.0\textwidth]{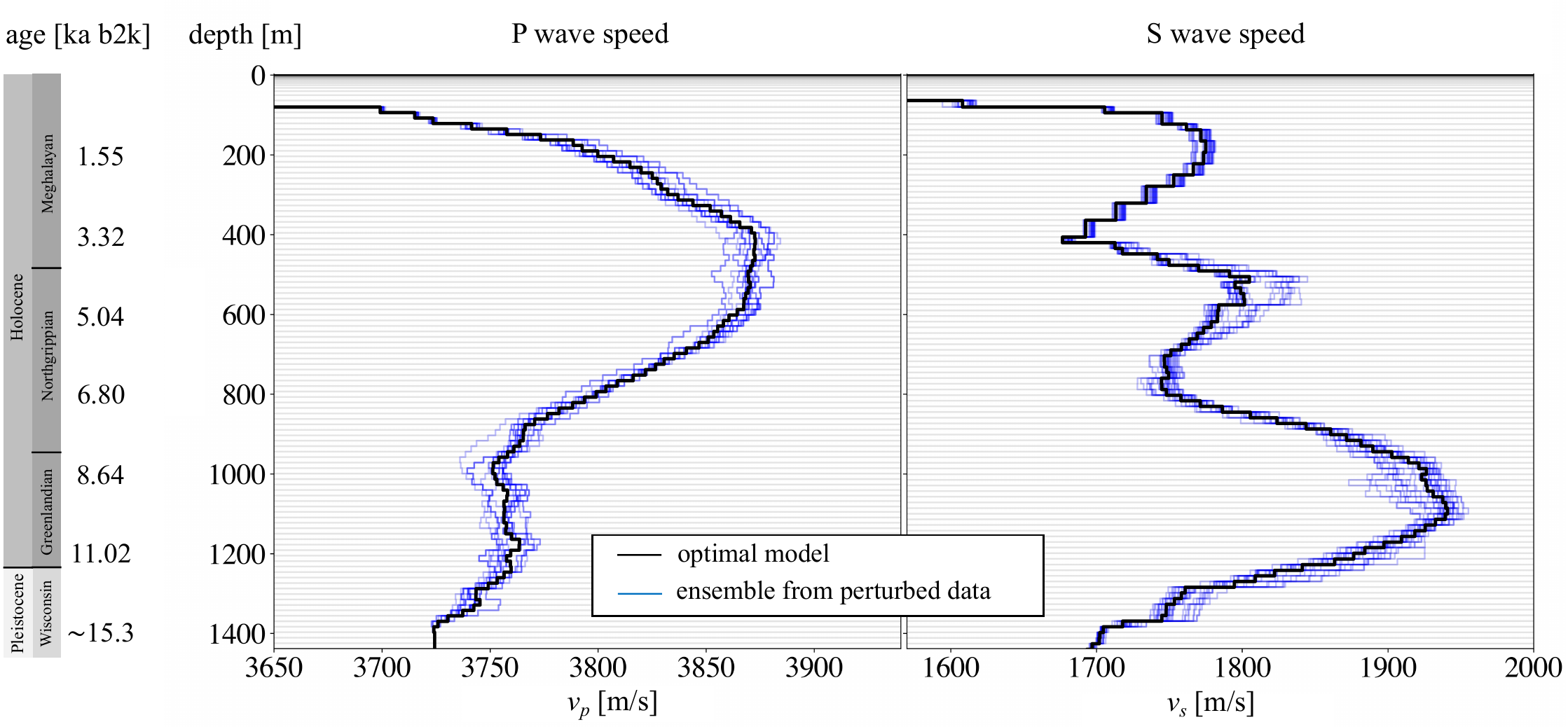}
\caption{Models of $v_p$ and $v_s$ as a function of depth. The black curves mark the optimal models. Models shown in blue represent the ensemble obtained by inverting traveltime data contaminated by random Gaussian errors with 1 ms standard deviation. For reference, the preliminary chronology to the left, in ka before 2000 (b2k) is taken from \cite{Mojtabavi_2020}.}
\label{F:velocity}
\end{center}
\end{figure}

Data from radar soundings on NEGIS, and around the EastGRIP camp in particular, constrain the topographic variations of layers within the ice stream to be on the order of only few metres over several kilometres distance \citep[e.g.,][]{Franke_2022,Mojtabavi_2022}.  This justifies an inversion of the traveltime observations for a stratified, laterally homogeneous medium.  To account for the dependence of ray paths on velocity structure,  we employ the iterative, nonlinear L-BFGS algorithm \citep[e.g.,][]{Nocedal_1980,Nocedal_1999,Fichtner_book_2021} for the minimisation of the $L_2$ misfit between observed and calculated traveltimes. The resulting models for $v_p$ and $v_s$ as a function of depth are shown in Fig. \ref{F:velocity}, together with a preliminary chronology of the EastGRIP ice core \citep{Mojtabavi_2020}.

\begin{figure}
\begin{center}
\noindent\includegraphics[width=0.9\textwidth]{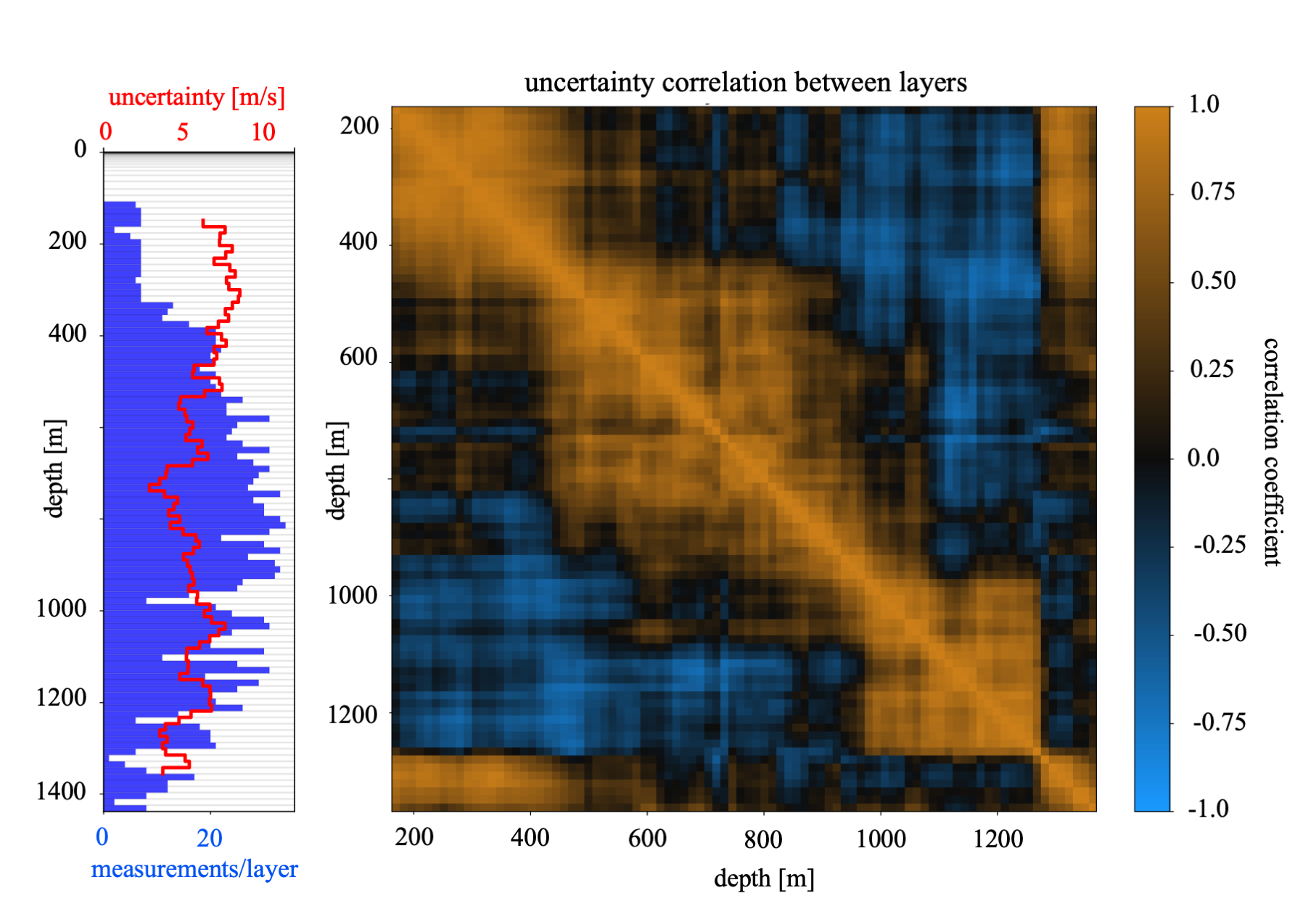}
\caption{Uncertainty proxies for the $v_p$ model. Model uncertainties correspond to the standard deviation of the $v_p$ ensemble shown in Fig. \ref{F:velocity}. They are shown only in the depth range where the L-BFGS optimisation produced significant deviations from the initial model, i.e., where gradients are significantly non-zero. The number of measurements per layer, in blue, is shown for comparison. The correlation matrix in the right panel is also derived from the $v_p$ ensemble. It describes the mapping of observational errors into correlated model uncertainties, i.e.,  the extent to which layers can be resolved individually.}
\label{F:errors_P}
\end{center}
\end{figure}

\begin{figure}
\begin{center}
\noindent\includegraphics[width=0.9\textwidth]{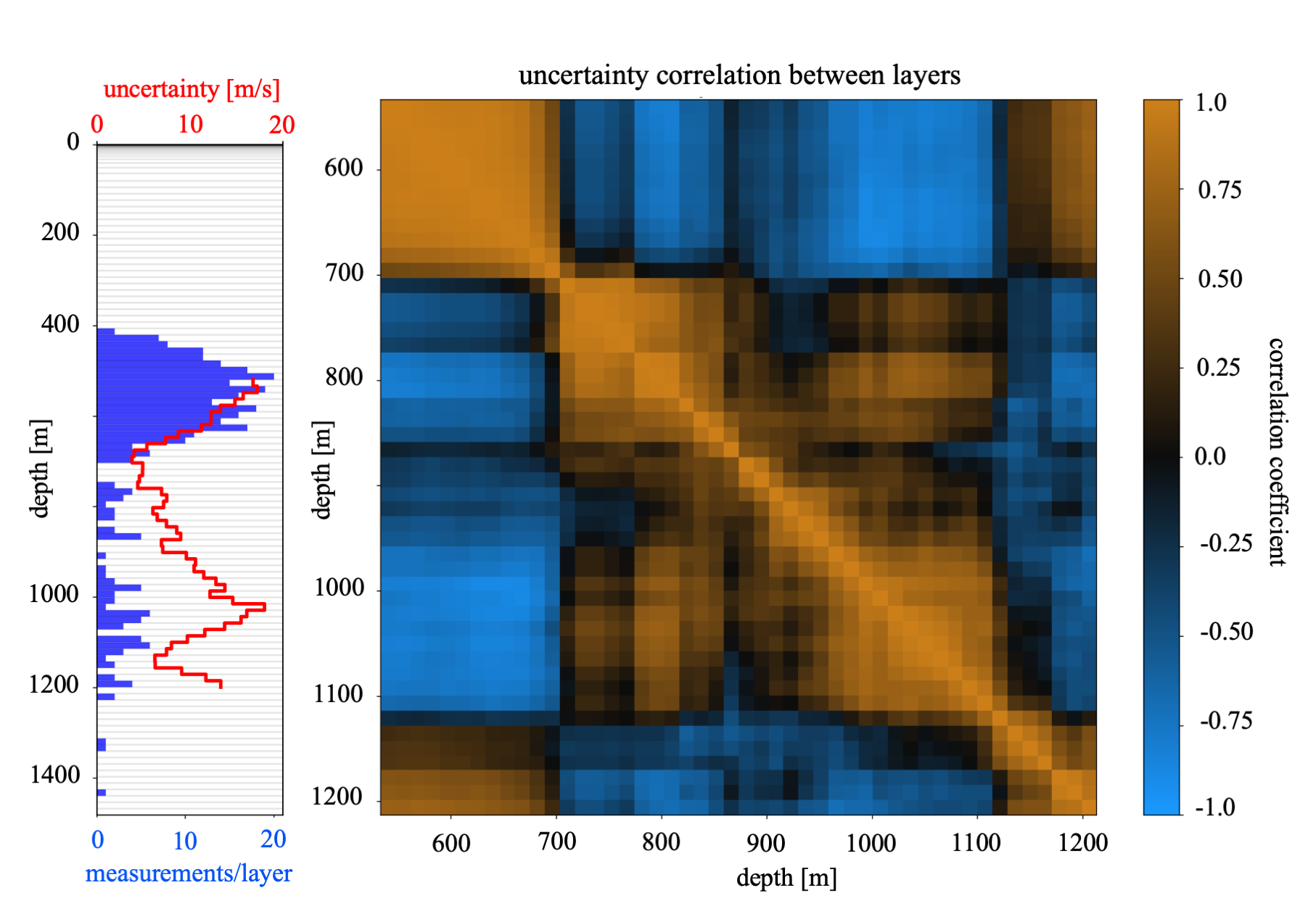}
\caption{The same as in Fig. \ref{F:errors_P} but for $v_s$ uncertainties.}
\label{F:errors_S}
\end{center}
\end{figure}

For an approximate assessment of model uncertainties induced by the combination of observational and forward modelling errors in the P and S wave traveltimes, we repeat the nonlinear inversions with datasets to which we added random Gaussian errors with 1 ms standard deviation. This results in the ensemble of 30 alternative $v_p$ and $v_s$ models shown as blue curves in Fig. \ref{F:velocity}. The ensembles themselves already provide a visual impression of model uncertainties. A more quantitative uncertainty analysis in the form of wave speed standard deviations and inter-layer correlations is displayed in Fig. \ref{F:errors_P} for $v_p$ and Fig. \ref{F:errors_S} for $v_s$.

\section{Reverse-time migration and reflectivity}\label{S:reflectivity}

Observations of upward reflected P waves, as shown in Fig. \ref{F:reduced}, complement the body wave traveltimes by providing information on structural details that are smaller than the dominant wavelength of $\sim$20 m. The spatial density of the DAS recordings facilitates the imaging of reflectivity in two ways.  First, reflected waves can be separated from the incident wave by simple $f$-$k$ filtering, in our case between velocities from $-4000$ to $-3600$ m$/$s, where the minus sign denotes upward propagation.  Second, a one-dimensional reverse-time migration can be computed for each channel individually; and the results can be averaged in order to reduce the effect of noise in the reflectivity image.

\begin{figure}
\begin{center}
\noindent\includegraphics[width=1.0\textwidth]{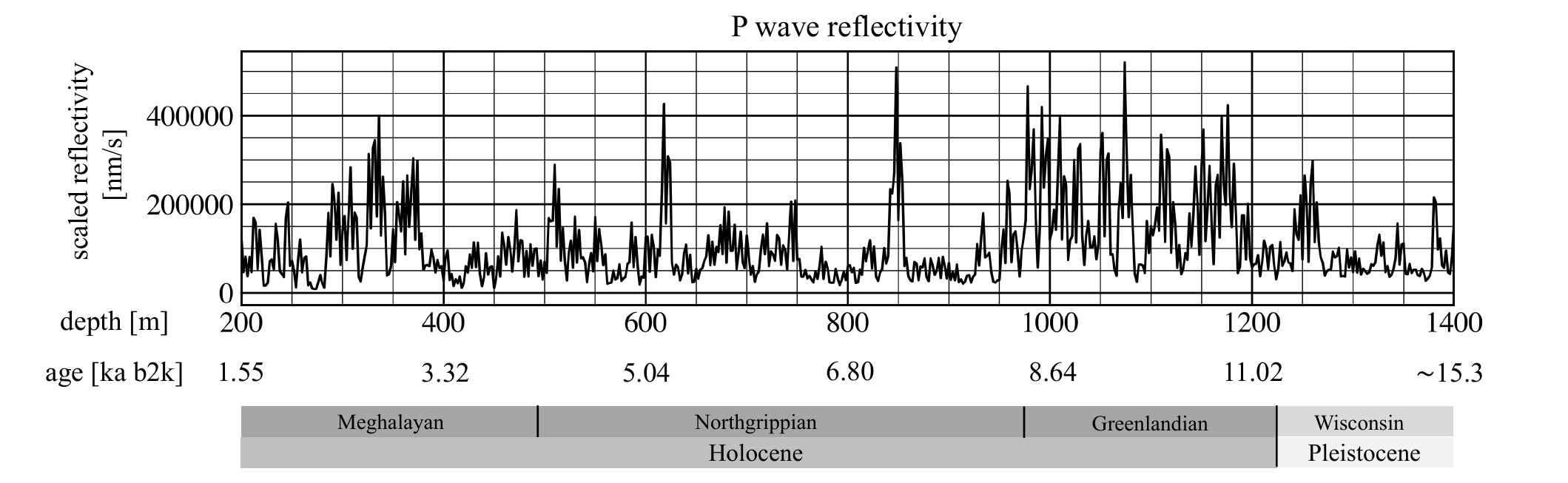}
\caption{P wave reflectivity as a function of depth, obtained  by reverse-time migration of the P wave reflections, summed over all channels between 200 and 1400 m depth. To compensate for geometric spreading, the results of the reverse-time migration at some depth $h$ are multiplied by $h$. This results in the physical unit nanostrain m$/$s = nm$/$s. The approximate age scale at the bottom is taken from \cite{Mojtabavi_2020}.}
\label{F:reflectivity}
\end{center}
\end{figure}

Fig. \ref{F:reflectivity} displays a scaled version of P wave reflectivity obtained by this procedure. To compensate for the amplitude loss due to geometric spreading, the value of the reflectivity image at some depth $h$ is multiplied by $h$. The reflectivity distribution features two pronounced peaks at 620 and 850 m depth.  As other, less prominent reflectivity peaks, they correspond to vertical discontinuities with a sharpness that is significantly smaller than the dominant P wavelength of $\sim$20 m. Furthermore, the depth intervals from $\sim$280 to $\sim$370 m and from $\sim$970 to $\sim$1180 m are characterised by a rapid sequence of high-contrast layers.

\section{Discussion}\label{S:Discussion}

The seismic wave speed models in Fig.  \ref{F:velocity} and the reflectivity image in Fig. \ref{F:reflectivity} carry information on the internal structure and the dynamics of the NEGIS in the vicinity of the EastGRIP drill site.  Examples include the variations of $v_p$ and $v_s$ in excess of 200 m$/$s. They suggest the presence of pronounced seismic anisotropy, which single ice crystal anisotropy suggests to be at approximately this level \citep[e.g.,][]{Diez_2015a}. Temperature variations that produce similar effects would have to reach $\sim$200 K \citep{Kohnen_1974}, nearly ten times of what is being observed in similarly deep boreholes on the Greenland Ice Sheet \citep{Lokkegaard_2022}, but only close to the ice-bed interface.  The more rapid changes of $v_s$ compared to $v_p$ hint at depth-dependent COF rotations predominantly in the horizontal plane that affect (quasi) S waves more strongly than mostly vertically propagating P waves \citep[e.g.,][]{Diez_2015a}.  Complementing the dynamic information contained in seismic anisotropy, the reflectivity image seems to carry a pronounced climatic imprint. The high reflectivity interval between $\sim$970 to $\sim$1180 m corresponds to the Greenlandian, the first stage of the Holocene, where higher temperatures caused an increased atmospheric dust load \citep{Fischer_2007} that leads to larger and more variable grain sizes in the ice \citep[e.g.,][]{Kerch_2016}.  A more quantitative interpretation of the fibre-optic sensing results requires a careful integration of complementary data, e.g., from ice core crytallography, borehole logging, and other surface-geophysical experiments, most notably radar sounding.  

Our focus here is on methodological aspects that pertain to fibre-optic seismology for deep ($>$1000 m) boreholes on glaciers and ice sheets in general. This includes imaging resolution and its limits, the derivation of other glaciologically relevant material properties from uncertain $v_p$ and $v_s$ profiles, and suggestions for improved experimental and analysis procedures.  We will discuss each of these aspects in the next subsection.

\subsection{Resolution and resolution limits}\label{SS:resolution}

One of the fundamental questions prior to our experiment concerned the achievable resolution, \emph{sensu lato}, of P and S wave speed variations, as well as its limiting factors. As shown in Fig. \ref{F:errors_P}, the large number of P wave traveltime measurements (1,896) constrains $v_p$ between $\sim$200 and $\sim$1400 m with an uncertainty of around 5 m$/$s or 0.14 \%, on average.  This uncertainty is roughly 40 times smaller than the $v_p$ variations within this depth interval.  In contrast, the number of S wave traveltime measurements is only 333,  resulting in significantly larger $v_s$ uncertainties of $\sim$10 m$/$s or 0.55 \% over the much smaller interval from 500--1200 m. 

These uncertainties are relative to the width of the layers in the velocity models, which we set to 15 m. This choice reflects the achievable spatial resolution, for which the uncertainty correlations in Figs. \ref{F:errors_P} and \ref{F:errors_S} serve as a useful proxy.  The half width of the inter-layer correlations ranges between $\sim$20 and $\sim$70 m, depending on depth. Hence, a layer thickness of 15 m ensures that resolvable features can just be represented by the model parameterisation.

Model uncertainties and correlation (resolution) lengths are primarily controlled by the combined measurement and modelling uncertainties.  The estimated $\sim$1 ms uncertainty of the traveltime measurements translates into propagation distances of around 2 m for S waves and 4 m for P waves.  Consequently, the selected 2 m channel spacing in our experiment is surely not among the limiting factors, unless measurement and modelling uncertainties can be reduced significantly.

A reduction of modelling uncertainties could be achieved through the replacement of ray theory by full-waveform modelling and inversion \citep[e.g.,][]{Fichtner_book,Virieux_2009,Liu_2012}, ideally taking anisotropy and visco-elastic attenuation into account.  However, the minimum wavelength of around 10 m for both P and S waves implies propagation distances 200 wavelengths or more, which is still challenging with currently available computational resources.

Major contributors to measurement uncertainties are SNR and signal bandwidth.  Nearly depth-independent  spectra in Fig. \ref{F:spectra}c suggest that the dominant noise is primarily of instrumental origin.  The extent to which future technological developments may reduce instrumental noise without compromising the spatial resolution of DAS measurements is hard to predict at this point.  Better coupling of the cable would almost certainly improve the SNR, but seems achievable without major efforts only by letting the cable freeze in after all other operations have seized, which is not an option for deep glaciological boreholes. 

Repeating the numerical experiment summarised in Fig. \ref{F:measurement_errors} with stretched and dilated wavelet templates, allows us to estimate the effect of signal bandwidth on the measurement error statistics. (Dilating a wavelet by a factor $n$, increases its frequency content by the same factor.) Owing to the approximate frequency independence of the noise between 100--500 Hz, a frequency doubling of the wavelet reduces the measurement uncertainties roughly by a factor of 2. This could be achieved with smaller detonation charges at the expense of poorer signal detection or by lowering charges into the borehole, which has obvious disadvantages.

While higher signal frequencies would also benefit the resolution of the reflectivity image, which is primarily wavelength-limited, it contradicts the need for stronger sources that increase the SNR and allow seismic energy to propagate above the noise level beyond 1500 m distance.

\subsection{Derived inferences}

\begin{figure}
\begin{center}
\noindent\includegraphics[width=0.9\textwidth]{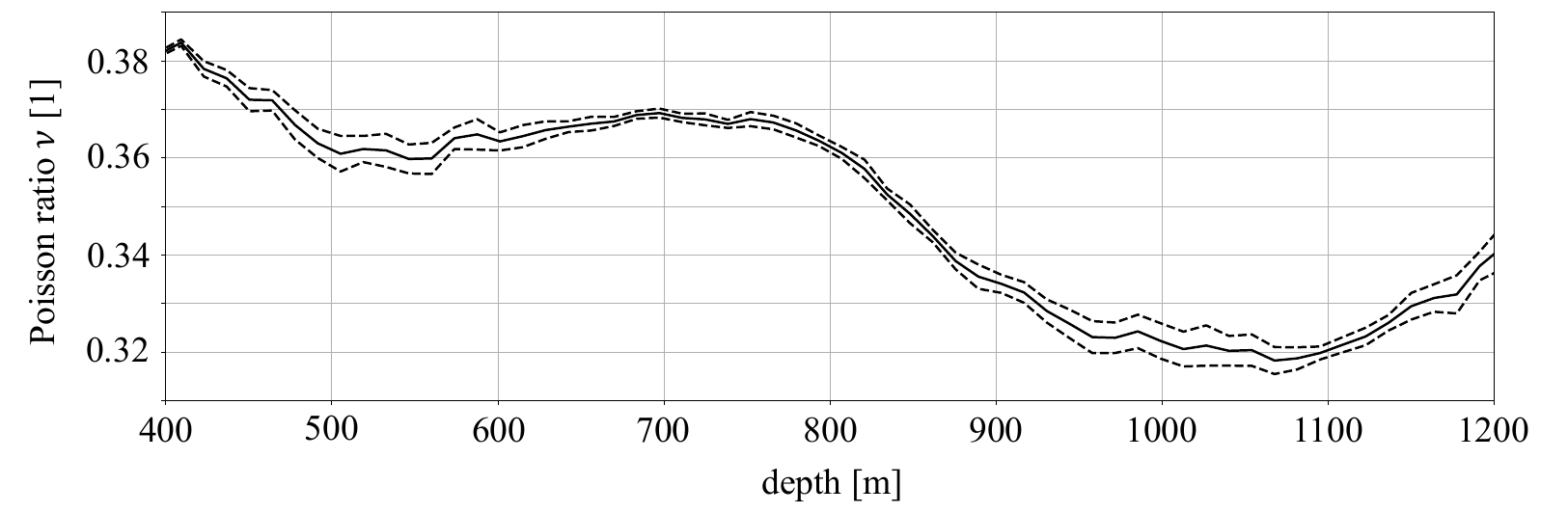}
\caption{Effective Poisson ratio $\nu$ within the depth range where both $v_p$ and $v_s$ can be constrained with reasonable confidence. Standard deviations from the mean are shown as dashed curves.}
\label{F:poisson}
\end{center}
\end{figure}

The joint inversion for $v_p$ and $v_s$ enables the estimation of derived quantities, such as the effective Poisson ratio
\begin{equation}\label{E:poisson}
\nu = \frac{v_p^2 - 2 v_s^2}{2(v_p^2-v_s^2)}\,.
\end{equation}
Since $v_p$ and $v_s$ are not equally well constrained at all depths,  $\nu$ and its standard deviation cannot be computed directly from (\ref{E:poisson}).  Instead, we produce Gaussian distributed random samples of $v_p$ and $v_s$, with means and standard deviations provided by the uncertainty estimates in Figs. \ref{F:errors_P} and \ref{F:errors_S}.  This results in a collection of samples for $\nu$, from which its mean and standard deviation can be computed. For the depth range where both $v_p$ and $v_s$ can be constrained with some confidence, $\nu$ is shown in Fig. \ref{F:poisson}.

While the procedure outlined above, illustrates how uncertainties in $v_p$ and $v_s$ can be used and propagated accurately into derived inferences,  it is important to note that the depth variations of the effective Poisson ratio $\nu$ carry the imprint of both anisotropy and the true Poisson ratio $\nu_0$, i.e., the deformation ratio in orthogonal directions.  Constraining $\nu_0$ would require a complete model of anisotropy as a function of depth.

Similar to $\nu$, other elastic parameters, such as the bulk or shear moduli may be constrained \citep[e.g.,][]{Schlegel_2019}. This would, however, require information about density, which is not directly accessible with seismic traveltime data. Though density can be estimated from seismic wave speeds using scaling relations \citep{Kohnen_1972,Diez_2014}, their uncertainties may be difficult to quantify.



\subsection{Logistics and future experiments}

In addition to any measures that reduce the data uncertainties listed in section \ref{SS:resolution}, future experiments may benefit from lowering the fibre-optic cable deeper into the borehole.  Especially in the deeper part of the ice sheets, where strong shear is present, changes in fabric and crystal size change the rheological properties considerably and determine the overall dynamics of the ice.  There is also a strong interest to record ice-bed events or other signals coming from fluid discharge that may help us to better understand the properties of the interface and the various processes that may affect the friction which is key for modelling. Our main limitation was the tensile strength, which could be improved by choosing a more sturdy cable.  The larger weight of such a cable would likely preclude any manual handling of the cable drums, and require not only a motorised winch but more sophisticated transportation solutions inside the underground drilling trench, as well as increased transport weights for remote deployment.

The precise effect of a higher tensile strength on the sensitivity to seismic wave motion in the borehole is hard to estimate \emph{a priori}. However, given that more tensile stress will be absorbed by the cable reinforcement, it is to be expected that sensitivity may decrease.

Irrespective of a particular cable type, sensitivity to deformation could be improved with better coupling. Fixing the cable behind a borehole casing is standard in seismic exploration but impractical in deep ice-core drilling, where a casing may only exist within the firn layer. One solution could be to let the cable freeze into the borehole after all other operations and \emph{in situ} analyses have seized. This would, however, require the non-trivial replacement of the drill fluid with water.  For shallower depth, hot-water drill boreholes may be an alternative.

\section{Conclusions}\label{S:Conclusions}

We presented initial results from the first DAS experiment in a deep ice-core borehole inside an active ice stream.  By far the most important conclusion at this stage is the logistic feasibility of such an experiment and the quality of the results.  The application of nonlinear traveltime tomography and the use of a firn model constrained by multi-mode surface wave data, allow us to invert for $v_p$ and $v_s$ with depth-dependent uncertainties on the order of only 10 m$/$s, and vertical resolution of 20--70 m. 

Although this resolution is on the same order of magnitude or even lower than that of standard crystallographic measurements along the ice core, the advantage of our approach is the sensing of a larger volume. Especially for deeper depths closer to the ice-bed interface this is important, as single crystals often have 10 cm in diameter and thus span the whole ice core and might thus bias ice-core based results. DAS-based results might thus yield results are more representative result for the bulk ice properties.

Thanks to the regularly spaced DAS data, upward reflections can be separated easily from the rest of the wavefield, and be used in a reverse-time migration that provides a detailed reflectivity image of the ice.  Constraining visco-elastic attenuation with reasonable confidence does not seem to be possible, because scattering is likely to be the dominant contributor to energy loss in the direct body waves.

With current levels of data space uncertainties, a spatial resolution of $v_p$ and $v_s$ on the order of the DAS channel spacing in the metre range is clearly not achievable.  Desirable improvements in this regard include (i) the simultaneous increase of the maximum frequency and the source strength, (ii) better coupling of the cable,  (iii) a reduction of instrumental noise, and (iv) the transition from ray-based to full-waveform modelling and inversion. While (i) contradicts itself to some extent, (ii) may only be achievable by letting the cable freeze in, i.e., by loosing both the borehole and the cable.  Realising that technological progress in the direction of (iii) is difficult to plan and predict, this leaves (iv) as the only option that can currently just be pulled off the shelf.

It follows, in summary, that significant improvements beyond the methods presented here will likely require a large jump in complexity and effort, with regards to both experimental and data analysis procedures. This, in turn, will require different project planning at all levels.


\begin{acknowledgments}
Andreas Fichtner and Coen Hofstede gratefully acknowledge support by the whole EastGRIP team, including the provision of all necessary infrastructure.  This experiment would not have been possible without the immense technical support by S{\o}ren B{\o}rsting, Sverrir Hilmarson and Dorthe-Dahl Jensen. Invaluable technical support, before and during the experiment, was provided by Silixa (Athena Chalari and support team) and Solifos (Andrea Fasciati). Olaf Eisen and Dimitri Zigone were supported by the CHIPSM grant of the University of Strasbourg Institute for Advanced Studies.  EastGRIP is directed and organized by the Centre for Ice and Climate at the Niels Bohr Institute, University of Copenhagen. It is supported by funding agencies and institutions in Denmark (A. P. Moller Foundation, University of Copenhagen), the United States (US National Science Foundation, Office of Polar Programs), Germany (Alfred Wegener Institute, Helmholtz Centre for Polar and Marine Research), Japan (National Institute of Polar Research and Arctic Challenge for Sustainability), Norway (University of Bergen and Trond Mohn Foundation), Switzerland (Swiss National Science Foundation), France (French Polar Institute Paul-Emile Victor, Institute for Geosciences and Environmental Research), Canada (University of Manitoba) and China (Chinese Academy of Sciences and Beijing Normal University).  
\end{acknowledgments}

\.\\
\noindent \textbf{Data statement}: All DAS data used in this work and the ensemble of final models are available on the Earth Model website of the ETH Seismology \& Wave Physics Group: www.swp.ethz.ch $\to$ Models.  Field work impressions, including the DAS cable deployment, can be found on the YouTube Channel of the ETH Seismology \& Wave Physics Group: https://www.youtube.com/@seismologyandwavephysics-e6406 .

\bibliography{biblio.bib}

\begin{thebibliography}{}

\bibitem[\protect\citeauthoryear{Aki and Richards}{Aki and
  Richards}{2002}]{Aki_Richards_2002}
Aki, K. and P.~Richards (2002).
\newblock {\em Quantitative Seismology}.
\newblock University Science Books.

\bibitem[\protect\citeauthoryear{Alley}{Alley}{1992}]{Alley_1992}
Alley, R.~B. (1992).
\newblock {Flow-law hypotheses for ice-sheet modeling}.
\newblock {\em J. Glaciol.\/}~{\em 129}, 245 -- 256.

\bibitem[\protect\citeauthoryear{Baig and Dahlen}{Baig and
  Dahlen}{2004}]{Baig_2004b}
Baig, A.~M. and F.~A. Dahlen (2004).
\newblock {Statistics of traveltimes and amplitudes in random media}.
\newblock {\em Geophys. J. Int.\/}~{\em 158}, 187--210.

\bibitem[\protect\citeauthoryear{Bennett}{Bennett}{1968}]{Bennett_1968}
Bennett, H.~F. (1968).
\newblock {\em An investigation into velocity anisotropy through measurements
  of ultrasonic wave velocities in snow and ice cores from Greenland and
  Antarctica}.
\newblock Ph.\ D. thesis, University of Wisconsin, Madison.

\bibitem[\protect\citeauthoryear{Bentley}{Bentley}{1972}]{Bentley_1972}
Bentley, C.~R. (1972).
\newblock {Seismic-wave velocities in anisotropic ice: A comparison of measured
  and calculated values in and around the deep drill hole at Byrd Stations,
  Antarctica}.
\newblock {\em J. Geophys. Res.\/}~{\em 77}, 4406--4420.

\bibitem[\protect\citeauthoryear{Bernauer, Behnen, Wassermann, Egdorf, Igel,
  Donner, Stammler, Hoffmann, Edme, Sollberger, Schmelzbach, Robertsson, Paitz,
  Igel, Smolinski, Fichtner, Rossi, Izgi, Vollmer, Eibl, Buske, Veress,
  Guattari, Laudat, Mattio, Sebe, Olivier, Lallemand, Brunner, Kurzych, Dudek,
  Jaroszewicz, Kowalski, Bonkowski, Bobra, Zembaty, Vackář, Málek, and
  Brokesova}{Bernauer et~al.}{2021}]{Bernauer_2021}
Bernauer, F., K.~Behnen, J.~Wassermann, S.~Egdorf, H.~Igel, S.~Donner,
  K.~Stammler, M.~Hoffmann, P.~Edme, D.~Sollberger, C.~Schmelzbach,
  J.~Robertsson, P.~Paitz, J.~Igel, K.~Smolinski, A.~Fichtner, Y.~Rossi,
  G.~Izgi, D.~Vollmer, E.~Eibl, S.~Buske, C.~Veress, F.~Guattari, T.~Laudat,
  L.~Mattio, O.~Sebe, S.~Olivier, C.~Lallemand, B.~Brunner, A.~Kurzych,
  M.~Dudek, L.~Jaroszewicz, J.~Kowalski, P.~Bonkowski, P.~Bobra, Z.~Zembaty,
  J.~Vackář, J.~Málek, and J.~Brokesova (2021).
\newblock {Rotation, Strain, and Translation Sensors Performance Tests with
  Active Seismic Sources}.
\newblock {\em Sensors\/}~{\em 21}, doi:10.3390/s21010264.

\bibitem[\protect\citeauthoryear{Blankenship and Bentley}{Blankenship and
  Bentley}{1987}]{Blankenship_1987}
Blankenship, D.~D. and C.~R. Bentley (1987).
\newblock {The crystalline fabric of polar ice sheets inferred from seismic
  anisotropy}.
\newblock {\em IAHS Publ.\/}~{\em 170}, 17--28.

\bibitem[\protect\citeauthoryear{Booth, Christoffersen, Schoonman, Clarke,
  Hubbard, Law, Doyle, Chudley, and Chalari}{Booth et~al.}{2020}]{Booth_2020}
Booth, A.~D., P.~Christoffersen, C.~Schoonman, A.~Clarke, B.~Hubbard, R.~Law,
  S.~H. Doyle, T.~R. Chudley, and A.~Chalari (2020).
\newblock {Distributed Acoustic Sensing of seismic properties in a borehole
  drilled on a fast-flowing Greenlandic outlet glacier}.
\newblock {\em Geophys. Res. Lett.\/}~{\em 47}, doi:10.1029/2020GL088148.

\bibitem[\protect\citeauthoryear{Brisbourne, Kendall, Kufner, Hudson, and
  Smith}{Brisbourne et~al.}{2021}]{Brisbourne_2021}
Brisbourne, A.~M., M.~Kendall, S.-K. Kufner, T.~S. Hudson, and A.~M. Smith
  (2021).
\newblock {Downhole distributed acoustic profiling at the Skytrain Ice Rise,
  West Antarctica}.
\newblock {\em The Cryosphere\/}~{\em 15}.

\bibitem[\protect\citeauthoryear{Brockamp and Pistor}{Brockamp and
  Pistor}{1967}]{Brockamp_1967}
Brockamp, B. and P.~Pistor (1967).
\newblock {Ein Beitrag zur seismischen Erforschung des Gr\"{ö}nl\"{a}ndischen
  Inlandeises}.
\newblock {\em Polarforschung\/}~{\em 37\/}(6), 133--146.

\bibitem[\protect\citeauthoryear{Chaves, Ritsema, and Koelemeijer}{Chaves
  et~al.}{2021}]{Chaves_2021}
Chaves, C. A.~M., J.~Ritsema, and P.~Koelemeijer (2021).
\newblock {Comparing ray-theoretical and finite-frequency teleseismic
  traveltimes: implications for constraining the ratio of S-wave to P-wave
  velocity variations in the lower mantle}.
\newblock {\em Geophys. J. Int.\/}~{\em 224}, 1540--1552.

\bibitem[\protect\citeauthoryear{Church, Clark, Cazenave, Gregory, Jevrejeva,
  Levermann, Merrifield, Milne, Nerem, Nunn, Payne, Pfeffer, Stammer, and
  Unnikrishnan}{Church et~al.}{2013}]{Church_2013}
Church, J.~A., P.~U. Clark, A.~Cazenave, J.~Gregory, S.~Jevrejeva,
  A.~Levermann, M.~Merrifield, G.~Milne, R.~Nerem, P.~Nunn, A.~Payne,
  W.~Pfeffer, D.~Stammer, and A.~Unnikrishnan (2013).
\newblock {Sea level change}.
\newblock In {\em Climate Change 2013: The Physical Science Basis. Contribution
  of Working Group I to the Fifth Assessment Report of the Intergovernmental
  Panel on Climate Change}. Cambridge University Press, Cambridge, UK.

\bibitem[\protect\citeauthoryear{Dahlen, Hung, and Nolet}{Dahlen
  et~al.}{2000}]{Dahlen_2000}
Dahlen, F., S.-H. Hung, and G.~Nolet (2000).
\newblock Fr\'{e}chet kernels for finite-frequency traveltimes -- {I}.
  {T}heory.
\newblock {\em Geophys. J. Int.\/}~{\em 141}, 157--174.

\bibitem[\protect\citeauthoryear{Daley, Pevzner, Shulakova, Kashikar, Miller,
  Goetz, and ans S.~Lueth}{Daley et~al.}{2013}]{Daley_2013}
Daley, T.~M., R.~Pevzner, V.~Shulakova, S.~Kashikar, D.~E. Miller, J.~Goetz,
  and J.~H. ans S.~Lueth (2013).
\newblock {Field testing of fiber-optic distributed acoustic sensing (DAS) for
  surbsurface seismic monitoring}.
\newblock {\em The Leading Edge\/}~{\em June 2013}, 936--942.

\bibitem[\protect\citeauthoryear{Daley, White, Miller, Robertson, Freifeld,
  Herkenhoff, and Cocker}{Daley et~al.}{2014}]{Daley_2014}
Daley, T.~M., D.~White, D.~E. Miller, M.~Robertson, B.~Freifeld, F.~Herkenhoff,
  and J.~Cocker (2014).
\newblock {Simultaneous acquisition of distributed acoustic sensing VSP with
  multi-mode and single-mode optical cables and 3-component geophones at the
  Aquistore CO$_2$ storage site}.
\newblock {\em SEG Extended Abstract\/}~{\em 2014}, 5014--5018.

\bibitem[\protect\citeauthoryear{Diez and Eisen}{Diez and
  Eisen}{2015}]{Diez_2015a}
Diez, A. and O.~Eisen (2015).
\newblock {Seismic wave propagation in anisotropic ice - Part 1: Elasticity
  tensor and derived quantities from ice-core properties}.
\newblock {\em The Cryosphere\/}~{\em 9}, 367--384.

\bibitem[\protect\citeauthoryear{Diez, Eisen, Hofstede, Lambrecht, Mayer,
  Miller, Steinhage, Binder, and Weikusat}{Diez et~al.}{2015}]{Diez_2015b}
Diez, A., O.~Eisen, C.~Hofstede, A.~Lambrecht, C.~Mayer, H.~Miller,
  D.~Steinhage, T.~Binder, and I.~Weikusat (2015).
\newblock {Seismic wave propagation in anisotropic ice - Part 2: Effects of
  crystal anisotropy in geophysical data}.
\newblock {\em The Cryosphere\/}~{\em 9}, 385--398.

\bibitem[\protect\citeauthoryear{Diez, Eisen, I.~Weikusat, Hofstede, Bohleber,
  Bohlen, and Polom}{Diez et~al.}{2014}]{Diez_2014}
Diez, A., O.~Eisen, J.~E. I.~Weikusat, C.~Hofstede, P.~Bohleber, T.~Bohlen, and
  U.~Polom (2014).
\newblock {Influence of ice crystal anisotropy on seismic velocity analysis}.
\newblock {\em Ann. Glaciol.\/}~{\em 55}, 97--106.

\bibitem[\protect\citeauthoryear{Diprinzio, Wilen, Alley, Fitzpatrick, Spencer,
  and Gow}{Diprinzio et~al.}{2005}]{Diprinzio_2005}
Diprinzio, C.~L., L.~A. Wilen, R.~B. Alley, J.~J. Fitzpatrick, M.~K. Spencer,
  and A.~J. Gow (2005).
\newblock {Fabric and texture at Siple Dome, Antarctica}.
\newblock {\em J. Glaciol.\/}~{\em 51}, 281--290.

\bibitem[\protect\citeauthoryear{Durand, Giller-Chaulet, Svensson, Gagliardini,
  Kipfstuhl, Meyssonnier, Parrenin, Duval, and Dahl-Jensen}{Durand
  et~al.}{2007}]{Durand_2007}
Durand, G., F.~Giller-Chaulet, A.~Svensson, O.~Gagliardini, S.~Kipfstuhl,
  J.~Meyssonnier, F.~Parrenin, P.~Duval, and D.~Dahl-Jensen (2007).
\newblock {Change in ice rheology during climate variations – implications
  for ice flow modelling and dating of the EPICA Dome C core}.
\newblock {\em Clim. Past\/}~{\em 3}, 155--167.

\bibitem[\protect\citeauthoryear{Fahnestock, Bindschadler, Kwok, and
  Jezek}{Fahnestock et~al.}{1993}]{Fahnestock_1993}
Fahnestock, M., R.~Bindschadler, R.~Kwok, and K.~Jezek (1993).
\newblock {Greenland Ice Sheet surface properties and ice dynamics from ERS-1
  SAR imagery}.
\newblock {\em Science\/}~{\em 262}, 1530--1534.

\bibitem[\protect\citeauthoryear{Faria, Weikusat, and Azuma}{Faria
  et~al.}{2013}]{Faria_2013}
Faria, S.~H., I.~Weikusat, and N.~Azuma (2013).
\newblock {The microstructure of polar ice. Part I: Highlights from ice core
  research}.
\newblock {\em J. Structural Geol.\/}~{\em 61}, 2--20.

\bibitem[\protect\citeauthoryear{Faria, Weikusat, and Azuma}{Faria
  et~al.}{2014}]{Faria_2014}
Faria, S.~H., I.~Weikusat, and N.~Azuma (2014).
\newblock {The microstructure of polar ice. Part II: State of the art}.
\newblock {\em J. Structural Geol.\/}~{\em 61}, 21--49.

\bibitem[\protect\citeauthoryear{Fichtner}{Fichtner}{2010}]{Fichtner_book}
Fichtner, A. (2010).
\newblock {\em Full {S}eismic {W}aveform {M}odelling and {I}nversion.}
\newblock Springer, Heidelberg.

\bibitem[\protect\citeauthoryear{Fichtner}{Fichtner}{2021}]{Fichtner_book_2021}
Fichtner, A. (2021).
\newblock {\em Lecture Notes on Inverse Theory}.
\newblock doi:10.33774/coe-2021-qpq2j: Cambridge Open Engage.

\bibitem[\protect\citeauthoryear{Fichtner, Hofstede, Kennett, Nymand,
  Lauritzen, Zigone, and Eisen}{Fichtner et~al.}{2023}]{Fichtner_2023}
Fichtner, A., C.~Hofstede, B.~L.~N. Kennett, N.~F. Nymand, M.~L. Lauritzen,
  D.~Zigone, and O.~Eisen (2023).
\newblock {Fiber-optic airplane seismology on the Northeast Greenalnd Ice
  Stream}.
\newblock {\em The Seismic Record\/}~{\em doi:10.1785/0320230004}, submitted.

\bibitem[\protect\citeauthoryear{Fichtner, Klaasen, Thrastarson, Cubuk-Sabuncu,
  Paitz, and Jonsdottir}{Fichtner et~al.}{2022}]{Fichtner_2022b}
Fichtner, A., S.~Klaasen, S.~Thrastarson, Y.~Cubuk-Sabuncu, P.~Paitz, and
  K.~Jonsdottir (2022).
\newblock {Fiber-optic observation of volcanic tremor through floating
  ice-sheet resonance}.
\newblock {\em The Seismic Record\/}~{\em 2}, 148--155.

\bibitem[\protect\citeauthoryear{Fischer, Siggaard-Andersen, Ruth,
  R\"{o}thlisberger, and Wolff}{Fischer et~al.}{2007}]{Fischer_2007}
Fischer, H., M.-L. Siggaard-Andersen, U.~Ruth, R.~R\"{o}thlisberger, and
  E.~Wolff (2007).
\newblock {Glacial/interglacial changes in mineral dust and sea-salt records in
  polar ice cores: Sources, transport, and deposition}.
\newblock {\em Rev. Geophys.\/}~{\em 45}, doi:10.1029/2005RG000192.

\bibitem[\protect\citeauthoryear{Franke, Jansen, Binder, Paden, D\"{o}rr,
  Gerber, Miller, {Dahl-Jensen}, Helm, Steinhage, Weikusat, Wilhelms, and
  Eisen}{Franke et~al.}{2022}]{Franke_2022}
Franke, S., D.~Jansen, T.~Binder, J.~D. Paden, N.~D\"{o}rr, T.~A. Gerber,
  H.~Miller, D.~{Dahl-Jensen}, V.~Helm, D.~Steinhage, I.~Weikusat, F.~Wilhelms,
  and O.~Eisen (2022).
\newblock {Airborne ultra-wideband radar sounding over the shear margins and
  along flow lines at the onset region of the Northeast Greenland Ice Stream}.
\newblock {\em Earth Syst. Sci. Data\/}~{\em 14}, 763--779.

\bibitem[\protect\citeauthoryear{Gerber, Lilien, Rathmann, Franke, Young,
  Valero-Delgado, Ershadi, Drews, Zeising, Humbert, Stoll, Weikusat, Grinsted,
  Hvidberg, Jansen, Miller, Helm, Steinhage, O'Neill, Paden, Gogineni,
  Dahl-Jensen, and Eisen}{Gerber et~al.}{2023}]{Gerber_2023}
Gerber, T.~A., D.~A. Lilien, N.~M. Rathmann, S.~Franke, T.~J. Young,
  F.~Valero-Delgado, M.~R. Ershadi, R.~Drews, O.~Zeising, A.~Humbert, N.~Stoll,
  I.~Weikusat, A.~Grinsted, C.~S. Hvidberg, D.~Jansen, H.~Miller, V.~Helm,
  D.~Steinhage, C.~O'Neill, J.~Paden, S.~P. Gogineni, D.~Dahl-Jensen, and
  O.~Eisen (2023).
\newblock {Crystal orientation fabric anisotropy causes directional hardening
  of the Northeast Greenland Ice Stream}.
\newblock {\em Nat. Comm.\/}~{\em 14}, doi:10.1038/s41467--023--38139--8.

\bibitem[\protect\citeauthoryear{Helm, Humbert, and Miller}{Helm
  et~al.}{2014}]{Helm_2014}
Helm, V., A.~Humbert, and H.~Miller (2014).
\newblock {Elevation and elevation change of Greenland and Antarctica derived
  from CryoSat-2}.
\newblock {\em The Cryosphere\/}~{\em 8}, 1539–1559.

\bibitem[\protect\citeauthoryear{Horgan, Anandakrishnan, Alley, Burkett, and
  Peters}{Horgan et~al.}{2011}]{Horgan_2011}
Horgan, H.~J., S.~Anandakrishnan, R.~B. Alley, P.~G. Burkett, and L.~E. Peters
  (2011).
\newblock {Englacial seismic reflectivity: imaging crystal orientation fabric
  in West Antarctica}.
\newblock {\em J. Glaciol.\/}~{\em 57}, 639--650.

\bibitem[\protect\citeauthoryear{Hudson, Baird, Kendall, Kufner, Brisbourne,
  Smith, Butcher, Chalari, and Clarke}{Hudson et~al.}{2021}]{Hudson_2021}
Hudson, T.~S., A.~F. Baird, J.~M. Kendall, S.~K. Kufner, A.~M. Brisbourne,
  A.~M. Smith, A.~Butcher, A.~Chalari, and A.~Clarke (2021).
\newblock {Distributed Acoustic Sensing (DAS) for natural microseismicity
  studies: A case study from Antarctica}.
\newblock {\em J. Geophys. Res.\/}~{\em 126}, doi:10.1029/2020JB021493.

\bibitem[\protect\citeauthoryear{Joughin, Smith, and Howat}{Joughin
  et~al.}{2018}]{Joughin_2018}
Joughin, I., B.~E. Smith, and I.~M. Howat (2018).
\newblock {A complete map of Greenland ice velocity derived from satellite data
  collected over 20 years}.
\newblock {\em J. Glaciol.\/}~{\em 64}, doi:10.1017/jog.2017.73.

\bibitem[\protect\citeauthoryear{Kennett}{Kennett}{2001}]{Kennett_2001}
Kennett, B. L.~N. (2001).
\newblock {\em The seismic wavefield I. - Introduction and theoretical
  development.}
\newblock Cambridge University Press.

\bibitem[\protect\citeauthoryear{Kerch}{Kerch}{2016}]{Kerch_2016}
Kerch, J.~K. (2016).
\newblock {\em Crystal-orientation fabric variations on the cm-scale in cold
  Alpine ice: Interaction with paleo-climate proxies under deformation and
  implications for the interpretation of seismic velocities}.
\newblock Doctoral thesis, Ruperto-Carola University of Heidelberg, Germany.

\bibitem[\protect\citeauthoryear{Khan, Choi, Morlighem, Rignot, Helm, Humbert,
  Mouginot, Millan, Kj{\ae}r, and Bj{\o}rk}{Khan et~al.}{2022}]{Khan_2022}
Khan, S.~A., Y.~Choi, M.~Morlighem, E.~Rignot, V.~Helm, A.~Humbert,
  J.~Mouginot, R.~Millan, K.~H. Kj{\ae}r, and A.~A. Bj{\o}rk (2022).
\newblock {Extensive inland thinning and speed-up of Northeast Greenland Ice
  Stream}.
\newblock {\em Nature\/}~{\em 611}, doi:10.1038/s41586--022--05301--z.

\bibitem[\protect\citeauthoryear{Khan, Kj{\ae}r, Bevis, Bamber, Wahr, Kjeldsen,
  Bj{\o}rk, Korsgaard, Stearns, {van den Broeke}, Liu, Larsen, and
  Muresan}{Khan et~al.}{2014}]{Khan_2014}
Khan, S.~A., K.~H. Kj{\ae}r, M.~Bevis, J.~L. Bamber, J.~Wahr, K.~K. Kjeldsen,
  A.~A. Bj{\o}rk, N.~J. Korsgaard, L.~A. Stearns, M.~R. {van den Broeke},
  L.~Liu, N.~K. Larsen, and I.~S. Muresan (2014).
\newblock {Sustained mass loss of the Northeast Greenland Ice Sheet triggered
  by regional warming}.
\newblock {\em Nat. Clim. Change\/}~{\em 4}, doi:10.1038/nclimate2161.

\bibitem[\protect\citeauthoryear{King, Howat, Candela, Noh, Jeong, No\"{e}l,
  {van den Broeke}, Wouters, and Negrete}{King et~al.}{2020}]{King_2020}
King, M.~D., I.~M. Howat, S.~G. Candela, M.~J. Noh, S.~Jeong, B.~P.~Y.
  No\"{e}l, M.~R. {van den Broeke}, B.~Wouters, and A.~Negrete (2020).
\newblock {Dynamic ice loss from the Greenland Ice Sheet driven by sustained
  glacier retreat}.
\newblock {\em Comm. Earth Env.\/}~{\em 1}, doi:10.1038/s43247--020--0001--2.

\bibitem[\protect\citeauthoryear{Klaasen, Paitz, Lindner, Dettmer, and
  Fichtner}{Klaasen et~al.}{2021}]{Klaasen_2021}
Klaasen, S., P.~Paitz, N.~Lindner, J.~Dettmer, and A.~Fichtner (2021).
\newblock {Distributed Acoustic Sensing in volcano-glacial environments —
  Mount Meager, British Columbia}.
\newblock {\em J. Geophys. Res.\/}~{\em 159}, doi:10.1029/2021JB022358.

\bibitem[\protect\citeauthoryear{Klaasen, Thrastarson, Fichtner, Cubuk-Sabuncu,
  and Jonsdottir}{Klaasen et~al.}{2022}]{Klaasen_2022}
Klaasen, S., S.~Thrastarson, A.~Fichtner, Y.~Cubuk-Sabuncu, and K.~Jonsdottir
  (2022).
\newblock {Sensing Iceland's most active volcano with a "buried hair"}.
\newblock {\em EOS\/}~{\em 103}, doi:10.1029/2022EO220007.

\bibitem[\protect\citeauthoryear{Kohnen}{Kohnen}{1972}]{Kohnen_1972}
Kohnen, H. (1972).
\newblock {\"{U}ber die Beziehung zwischen seismischen Geschwindigkeiten und
  der Dichte in Firn und Eis}.
\newblock {\em Zeitschrift f. Geophysik\/}, 925--935.

\bibitem[\protect\citeauthoryear{Kohnen}{Kohnen}{1974}]{Kohnen_1974}
Kohnen, H. (1974).
\newblock {The temperature dependence of seismic waves in ice}.
\newblock {\em J. Glaciology\/}, 144--147.

\bibitem[\protect\citeauthoryear{Kohnen and Bentley}{Kohnen and
  Bentley}{1973}]{Kohnen_1973}
Kohnen, H. and C.~R. Bentley (1973).
\newblock {Seismic refraction and reflection measurements at Byrd Station,
  Antarctica}.
\newblock {\em J. Glaciology\/}, 101--111.

\bibitem[\protect\citeauthoryear{Lindsey, Rademacher, and
  {Ajo-Franklin}}{Lindsey et~al.}{2020}]{Lindsey_2020}
Lindsey, N.~J., H.~Rademacher, and J.~B. {Ajo-Franklin} (2020).
\newblock On the broadband instrument response of fiber-optic {DAS} arrays.
\newblock {\em J. Geophys. Res.\/}~{\em 125}, doi.org:10.1029/2019JB018145.

\bibitem[\protect\citeauthoryear{Liu and Gu}{Liu and Gu}{2012}]{Liu_2012}
Liu, Q. and Y.~Gu (2012).
\newblock {Seismic imaging: from classical to adjoint tomography}.
\newblock {\em Tectonophysics\/}~{\em 566-567}, 31--66.

\bibitem[\protect\citeauthoryear{L{\o}kkegaard, Mankoff, Zdanowicz, Clow,
  L\"uthi, Doyle, Thomsen, Fisher, Harper, Aschwanden, Vinther, Dahl-Jensen,
  Zekollari, Meierbachtol, McDowell, Humphrey, Solgaard, Karlsson, Khan, Hills,
  Law, Hubbard, Christoffersen, Jacquemart, Fausto, and Colgan}{L{\o}kkegaard
  et~al.}{2022}]{Lokkegaard_2022}
L{\o}kkegaard, A., K.~Mankoff, C.~Zdanowicz, G.~D. Clow, M.~P. L\"uthi,
  S.~Doyle, H.~Thomsen, D.~Fisher, J.~Harper, A.~Aschwanden, B.~M. Vinther,
  D.~Dahl-Jensen, H.~Zekollari, T.~Meierbachtol, I.~McDowell, N.~Humphrey,
  A.~Solgaard, N.~B. Karlsson, S.~A. Khan, B.~Hills, R.~Law, B.~Hubbard,
  P.~Christoffersen, M.~Jacquemart, R.~S. Fausto, and W.~T. Colgan (2022).
\newblock Greenland and canadian arctic ice temperature profiles.
\newblock {\em The Cryosphere Discussions\/}~{\em 2022}, 1--24.

\bibitem[\protect\citeauthoryear{Luo and Schuster}{Luo and
  Schuster}{1991}]{Luo_Schuster_1991}
Luo, Y. and G.~T. Schuster (1991).
\newblock Wave-equation traveltime inversion.
\newblock {\em Geophysics\/}~{\em 56}, 645--653.

\bibitem[\protect\citeauthoryear{Mankoff, Colgan, Solgaard, Karlsson, m, {van
  As}, Box, Khan, Kjeldsen, and Fausto}{Mankoff et~al.}{2019}]{Mankoff_2019}
Mankoff, K.~D., W.~Colgan, A.~Solgaard, N.~B. Karlsson, A.~P.~A. m, D.~{van
  As}, J.~E. Box, S.~A. Khan, K.~K. Kjeldsen, and J.~M. R.~S. Fausto (2019).
\newblock {Greenland Ice Sheet solid ice discharge from 1986 through 2017}.
\newblock {\em Earth Sys. Sci. Data\/}~{\em 11}, 769--786.

\bibitem[\protect\citeauthoryear{Marquering, Dahlen, and Nolet}{Marquering
  et~al.}{1999}]{Marquering_1999}
Marquering, H., F.~A. Dahlen, and G.~Nolet (1999).
\newblock {Three-dimensional sensitivity kernels for finite-frequency
  traveltimes: the banana-doughnut paradox}.
\newblock {\em Geophys. J. Int.\/}~{\em 137}, 805--815.

\bibitem[\protect\citeauthoryear{Mart\'{i}n, Gudmundsson, Pritchard, and
  Gagliardini}{Mart\'{i}n et~al.}{2009}]{Martin_2009}
Mart\'{i}n, C., G.~H. Gudmundsson, H.~D. Pritchard, and O.~Gagliardini (2009).
\newblock {On the effects of anisotropic rheology on ice flow, internal
  structure, and the age-depth relationship at ice divides}.
\newblock {\em J. Geophys. Res.\/}~{\em 114}, doi:10.1029/2008JF001204.

\bibitem[\protect\citeauthoryear{Mateeva, Lopez, Mestayer, Wills, Cox,
  Kiyashchenko, Yang, Berlang, Detomo, and Grandi}{Mateeva
  et~al.}{2013}]{Mateeva_2013}
Mateeva, A., J.~Lopez, J.~Mestayer, P.~Wills, B.~Cox, D.~Kiyashchenko, Z.~Yang,
  W.~Berlang, R.~Detomo, and S.~Grandi (2013).
\newblock Distributed acoustic sensing for reservoir monitoring with {VSP}.
\newblock {\em The Leading Edge\/}~{\em October 2013}, 1278--1283.

\bibitem[\protect\citeauthoryear{Mateeva, Lopez, Potters, Mestayer, Cox,
  Kiyashchenko, Wills, Grandi, Kuvshinov, Berlang, Yang, and Detomo}{Mateeva
  et~al.}{2014}]{Mateeva_2014}
Mateeva, A., J.~Lopez, H.~Potters, J.~Mestayer, B.~Cox, D.~Kiyashchenko,
  P.~Wills, S.~Grandi, B.~Kuvshinov, W.~Berlang, Z.~Yang, and R.~Detomo (2014).
\newblock {Distributed acoustic sensing for reservoir monitoring with vertical
  seismic profiling}.
\newblock {\em Geophys. Prosp.\/}~{\em 62}, 679--692.

\bibitem[\protect\citeauthoryear{Mojtabavi, Eisen, Franke, Jansen, Steinhage,
  Paden, {Dahl-Jensen}, Weikusat, Eichler, and Wilhelms}{Mojtabavi
  et~al.}{2022}]{Mojtabavi_2022}
Mojtabavi, S., O.~Eisen, S.~Franke, D.~Jansen, D.~Steinhage, J.~Paden,
  D.~{Dahl-Jensen}, I.~Weikusat, J.~Eichler, and F.~Wilhelms (2022).
\newblock {Origin of englacial stratigraphy at three deep ice core sites of the
  Greenland Ice Sheet by synthetic radar modelling}.
\newblock {\em J. Glac.\/}~{\em 68}, 799--811.

\bibitem[\protect\citeauthoryear{Mojtabavi, Wilhelms, Cook, Davies, Sinnl,
  Jensen, Dahl-Jensen, Svensson, Vinther, Kipfstuhl, Jones, Karlsson, Faria,
  Gkinis, Kjaer, Erhardt, Berben, Nisancioglu, Koldtoft, and
  Rasmussen}{Mojtabavi et~al.}{2020}]{Mojtabavi_2020}
Mojtabavi, S., F.~Wilhelms, E.~Cook, S.~M. Davies, G.~Sinnl, M.~S. Jensen,
  D.~Dahl-Jensen, A.~Svensson, B.~M. Vinther, S.~Kipfstuhl, G.~Jones, N.~B.
  Karlsson, S.~H. Faria, V.~Gkinis, H.~A. Kjaer, T.~Erhardt, S.~M.~P. Berben,
  K.~H. Nisancioglu, I.~Koldtoft, and S.~O. Rasmussen (2020).
\newblock {A first chronology for the East Greenland Ice-core Project (EGRIP)
  over the Holocene and last glacial termination}.
\newblock {\em Clim. Past\/}~{\em 16}, doi:10.5194/cp--16--2359--2020.

\bibitem[\protect\citeauthoryear{Mouginot, Rigot, rk, {van den Broeke}, Millan,
  Morlighem, No\"{e}l, Scheuchl, and Wood}{Mouginot
  et~al.}{2019}]{Mouginot_2019}
Mouginot, J., E.~Rigot, A.~A.~B. rk, M.~{van den Broeke}, R.~Millan,
  M.~Morlighem, B.~No\"{e}l, B.~Scheuchl, and M.~Wood (2019).
\newblock {Forty-six years of Greenland Ice Sheet mass balance from 1972 -
  2018}.
\newblock {\em Proc. Nat. Acad. Sci. USA\/}~{\em 116}, 9239--9244.

\bibitem[\protect\citeauthoryear{Nocedal}{Nocedal}{1980}]{Nocedal_1980}
Nocedal, J. (1980).
\newblock {Updating quasi-Newton matrices with limited storage}.
\newblock {\em Math. Comp.\/}~{\em 35}, 773--782.

\bibitem[\protect\citeauthoryear{Nocedal and Wright}{Nocedal and
  Wright}{1999}]{Nocedal_1999}
Nocedal, J. and S.~J. Wright (1999).
\newblock {\em Numerical Optimization}.
\newblock Springer, New York.

\bibitem[\protect\citeauthoryear{Paitz, Edme, Gr\"{a}ff, Walter, Doetsch,
  Chalari, Schmelzbach, and Fichtner}{Paitz et~al.}{2021}]{Paitz_2021}
Paitz, P., P.~Edme, D.~Gr\"{a}ff, F.~Walter, J.~Doetsch, A.~Chalari,
  C.~Schmelzbach, and A.~Fichtner (2021).
\newblock Empirical investigations of the instrument response for distributed
  acoustic sensing ({DAS}) across 17 octaves.
\newblock {\em Bull. Seis. Soc. Am.\/}~{\em 111}, 1--10.

\bibitem[\protect\citeauthoryear{Pettit, Thorsteinsson, Jacobson, and
  Waddington}{Pettit et~al.}{2007}]{Pettit_2007}
Pettit, E.~C., T.~Thorsteinsson, H.~P. Jacobson, and E.~D. Waddington (2007).
\newblock {The role of crystal fabric in flow near an ice divide}.
\newblock {\em J. Glaciol.\/}~{\em 53}, 277--288.

\bibitem[\protect\citeauthoryear{Picotti, Vuan, Carcione, Horgan, and
  Anandakrishnan}{Picotti et~al.}{2015}]{Picotti_2015}
Picotti, S., A.~Vuan, J.~M. Carcione, H.~J. Horgan, and S.~Anandakrishnan
  (2015).
\newblock {Anisotropy and crystalline fabric of Whillans Ice Stream (West
  Antarctica) inferred from multicomponent seismic data}.
\newblock {\em J. Geophys. Res.\/}~{\em 120}, 4237--4262.

\bibitem[\protect\citeauthoryear{Rignot and Mouginot}{Rignot and
  Mouginot}{2012}]{Rignot_2012}
Rignot, E. and J.~Mouginot (2012).
\newblock {Ice flow in Greenland for the international polar year 2008-2009}.
\newblock {\em Geophys. Res. Lett.\/}~{\em 39}, doi:10.1029/2012GL051634.

\bibitem[\protect\citeauthoryear{Samyn, Svensson, Fitzsimins, and
  Lorrain}{Samyn et~al.}{2005}]{Samyn_2005}
Samyn, D., A.~Svensson, S.~J. Fitzsimins, and R.~D. Lorrain (2005).
\newblock {Ice crystal properties of amber ice and strain enhancement at the
  base of cold Antarctic glaciers}.
\newblock {\em Ann. Glaciol.\/}~{\em 40}, 185--190.

\bibitem[\protect\citeauthoryear{Schlegel, Diez, L\"{o}we, Mayer, Lambrecht,
  Freitag, Miller, Hofstede, and Eisen}{Schlegel et~al.}{2019}]{Schlegel_2019}
Schlegel, R., A.~Diez, H.~L\"{o}we, C.~Mayer, A.~Lambrecht, J.~Freitag,
  H.~Miller, C.~Hofstede, and O.~Eisen (2019).
\newblock Comparison of elastic moduli from seismic diving-wave and ice-core
  microstructure analysis in antarctic polar firn.
\newblock {\em Ann. Glaciol.\/}~{\em 60}, 220--230.

\bibitem[\protect\citeauthoryear{Sheldon, Steffensen, Hansen, Popp, and
  Johnsen}{Sheldon et~al.}{2014}]{Sheldon_2014}
Sheldon, S.~G., J.~P. Steffensen, S.~B. Hansen, T.~J. Popp, and S.~J. Johnsen
  (2014).
\newblock {The investigation and experience of using ESTISOL\texttrademark 240
  and COASOL\texttrademark for ice-core drilling}.
\newblock {\em Ann. Glac.\/}~{\em 55}, 219--232.

\bibitem[\protect\citeauthoryear{Turin}{Turin}{1960}]{Turin_1960}
Turin, G.~L. (1960).
\newblock An introduction to matched filters.
\newblock {\em IRE Trans. Inf. Theo.\/}~{\em 6}, 311--329.

\bibitem[\protect\citeauthoryear{Vallelonga, Christianson, Alley,
  Anandakrishnan, Christian, Dahl-Jensen, Gkinis, Holme, Jacobel, Karlsson,
  Keisling, Kipfstuhl, Kj{\ae}r, Kristensen, Muto, Peters, Popp, Riverman,
  Svensson, Tibuleac, Vinther, Weng, and Winstrup}{Vallelonga
  et~al.}{2014}]{Vallelonga_2014}
Vallelonga, P., K.~Christianson, R.~B. Alley, S.~Anandakrishnan, J.~E.~M.
  Christian, D.~Dahl-Jensen, V.~Gkinis, C.~Holme, R.~W. Jacobel, N.~B.
  Karlsson, B.~A. Keisling, S.~Kipfstuhl, H.~A. Kj{\ae}r, M.~E.~L. Kristensen,
  A.~Muto, L.~E. Peters, T.~Popp, K.~L. Riverman, A.~M. Svensson, C.~Tibuleac,
  B.~M. Vinther, Y.~Weng, and M.~Winstrup (2014).
\newblock {Initial results from geophysical surveys and shallow coring of the
  Northeast Greenland Ice Stream (BEGIS)}.
\newblock {\em The Cryosphere\/}~{\em 8}, 1275--1287.

\bibitem[\protect\citeauthoryear{Virieux and Operto}{Virieux and
  Operto}{2009}]{Virieux_2009}
Virieux, J. and S.~Operto (2009).
\newblock An overview of full waveform inversion in exploration geophysics.
\newblock {\em Geophysics\/}~{\em 74}, WCC127--WCC152.

\bibitem[\protect\citeauthoryear{Walter, Gr\"{a}ff, Lindner, Paitz, K\"{o}pfli,
  Chmiel, and Fichtner}{Walter et~al.}{2020}]{Walter_2020}
Walter, F., D.~Gr\"{a}ff, F.~Lindner, P.~Paitz, M.~K\"{o}pfli, M.~Chmiel, and
  A.~Fichtner (2020).
\newblock {Distributed Acoustic Sensing of microseismic sources and wave
  propagation in glaciated terrain}.
\newblock {\em Nat. Comm.\/}~{\em 11}, doi:10.1038/s41467--020--15824.

\bibitem[\protect\citeauthoryear{Zhou, Butcher, Brisbourne, Kufner, Kendall,
  and Stork}{Zhou et~al.}{2023}]{Zhou_2023}
Zhou, W., A.~Butcher, A.~Brisbourne, S.-K. Kufner, J.-M. Kendall, and A.~Stork
  (2023).
\newblock {Seismic noise interferometry and Distributed Acoustic Sensing (DAS):
  measuring the firn layer S-velocity structure on Rutford Ice Stream,
  Antarctica}.
\newblock {\em J. Geophys. Res.\/}~{\em 127}, doi:10.1029/2022JF006917.

\end{thebibliography}
 
\end{document}